\documentclass[letter, 11pt]{article}
\pdfoutput=1

\usepackage{mathtools}
\usepackage{amssymb}
\usepackage[a4paper]{geometry}
\usepackage{graphicx}
\usepackage{microtype}
\usepackage{siunitx}
\usepackage{booktabs}
\usepackage[colorlinks=false, pdfborder={0 0 0}]{hyperref}
\usepackage{cleveref}
\usepackage{verbatim}
\usepackage[center,small,it]{caption}
\usepackage{multirow}
\usepackage{longtable}
\usepackage{tabu}
\usepackage{amsfonts}
\usepackage{tikz}
\usepackage[square]{natbib}
\usepackage{times}

\newcolumntype{L}{>{\ttfamily\raggedright}l} \newcolumntype{J}{>{\ttfamily\itshape\raggedright}l}

\makeatletter
\pgfdeclareshape{slash underlined}
{
  \inheritsavedanchors[from=rectangle]   \inheritanchorborder[from=rectangle]
  \inheritanchor[from=rectangle]{north}
  \inheritanchor[from=rectangle]{north west}
  \inheritanchor[from=rectangle]{north east}
  \inheritanchor[from=rectangle]{center}
  \inheritanchor[from=rectangle]{west}
  \inheritanchor[from=rectangle]{east}
  \inheritanchor[from=rectangle]{mid}
  \inheritanchor[from=rectangle]{mid west}
  \inheritanchor[from=rectangle]{mid east}
  \inheritanchor[from=rectangle]{base}
  \inheritanchor[from=rectangle]{base west}
  \inheritanchor[from=rectangle]{base east}
  \inheritanchor[from=rectangle]{south}
  \inheritanchor[from=rectangle]{south west}
  \inheritanchor[from=rectangle]{south east}
  \inheritanchorborder[from=rectangle]
  \foregroundpath{
    \southwest \pgf@xa=\pgf@x \pgf@ya=\pgf@y
    \northeast \pgf@xb=\pgf@x \pgf@yb=\pgf@y
    \pgf@xc=\pgf@xa
    \advance\pgf@xc by .5\pgf@xb
    \pgf@yc=\pgf@ya
    \advance\pgf@xc by -1.3pt
    \advance\pgf@yc by -1.8pt
    \pgfpathmoveto{\pgfqpoint{\pgf@xc}{\pgf@yc}}
    \advance\pgf@xc by  2.6pt
    \advance\pgf@yc by  3.6pt
    \pgfpathlineto{\pgfqpoint{\pgf@xc}{\pgf@yc}}
    \pgfpathmoveto{\pgfqpoint{\pgf@xa}{\pgf@ya}}
    \pgfpathlineto{\pgfqpoint{\pgf@xb}{\pgf@ya}}
 }
}
\tikzset{nomorepostaction/.code=\let\tikz@postactions\pgfutil@empty}
\newcommand\xnrightarrow[2][]{  \mathrel{\tikz[baseline=-.4ex] \path node[slash underlined,draw,->,anchor=south] {\(\scriptstyle #2\)} node[anchor=north] {\(\scriptstyle #1\)};}}
\makeatother

\newcommand{\RVAL}{\textsc{RVal}(\mathcal{F})}
\mathchardef\mhyphen="2D

\newcommand{\citeN}{\citet}
\renewcommand{\cite}{\citep}

\newcommand{\A}{\ensuremath{A}}

\begin{document}

\title{Consistency in Non-Transactional \\ Distributed Storage Systems}

\author{
  Paolo Viotti\\
  EURECOM \\
  \texttt{\href{mailto:viotti@eurecom.fr}{viotti@eurecom.fr}}
  \and
  Marko Vukoli\'c\\
  IBM Research - Zurich \\
  \texttt{\href{mailto:mvu@zurich.ibm.com}{mvu@zurich.ibm.com}}
}

\date{}

\maketitle
\thispagestyle{empty}
\begin{abstract}
Over the years, different meanings have been associated to the word \emph{consistency}
in the distributed systems community.  While in the '80s ``consistency'' typically meant \emph{strong consistency}, later defined also as \emph{linearizability}, in recent years, with the advent of highly available and scalable systems, the notion of ``consistency'' has been at the same time both weakened and blurred. 

In this paper we aim to fill the void in literature, by providing a structured and comprehensive overview of different consistency notions that appeared in distributed systems, and in particular \emph{storage} systems research, in the last four decades. We overview more than 50 different consistency notions, ranging from linearizability to eventual and weak consistency, defining precisely many of these, in particular where the previous definitions were ambiguous. We further provide a partial order among different consistency predicates, ordering them by their semantic ``strength'', which we believe will reveal useful in future research. Finally, we map the consistency semantics to different practical systems and research prototypes.

The scope of this paper is restricted to non-transactional semantics, i.e., those that apply to single storage object operations. As such, our paper complements the existing surveys done in the context of transactional, database consistency semantics. 
\end{abstract}

\newpage
\setcounter{page}{1}

\section{Introduction}

Faced with the inherent challenges of failures, communication/computation asynchrony and concurrent access to shared resources, distributed system designers have continuously sought to hide these fundamental concerns from users by offering abstractions and semantic models of various strength. At first glance, the ultimate goal of a distributed system is seemingly simple, as it should ideally be just a fault-tolerant and more scalable version of a centralized system. Namely, an ideal distributed system should leverage distribution and replication to boost availability by masking failures, provide scalability and/or reduce latency, but maintain the simplicity of use of a centralized system -- and, notably, its \emph{consistency} -- providing the illusion of sequential access. Such \emph{strong} consistency criteria can be found in early seminal works that paved the way of modern storage systems, e.g., \cite{Lamport:78,Lamport:86:vol1}, as well as  
in the subsequent advances in defining general, practical correctness conditions, such as \emph{linearizability} \cite{Herlihy.Wing:90}.

Unfortunately, the goals of high availability and strong consistency, in particular  linearizability, have been identified as mutually conflicting in many practical circumstances. Negative theoretical results and lower bounds, such as the FLP impossibility result \cite{Fischer.ea:85} and the CAP theorem \cite{Gilbert.Lynch:02}, shaped the design space of distributed systems. As a result, distributed system designers have either to give up the idealized goals of scalability and availability, or relax consistency. 

In recent years, the rise of commercial Internet-scale wide area computing caused system designers to prefer availability over consistency, leading to the advent of weak and eventual consistency \cite{Terry.Demers.ea:94,Saito.Shapiro:05,Vogels:08}.
Consequently, much research has been focusing on attaining a better understanding of
those weaker semantics \cite{Bailis.Ghodsi:13}, but also on adapting \cite{Bailis.Fakete.ea:14} 
or dismissing and replacing stronger ones \cite{Helland:07}.
Along this line of research, tools have been conceived in order to deal with consistency at the level of programming languages \cite{Alvaro.ea:11},
data objects \cite{Shapiro.ea:11,Burckhardt.ea:12} or data flows \cite{Alvaro.ea:13}.

Today, however, after roughly four decades of intensive and exciting research on various flavors of consistency, we lack a structured and comprehensive overview of different consistency notions that appeared in distributed systems research, and \emph{storage} systems research, in particular. 

This paper aims to help fill this void, by giving an overview of over 50 different consistency notions, ranging from linearizability to eventual and weak consistency, defining precisely many of these, in particular where the previous definitions were ambiguous. We further provide a partial order among different consistency notions, ordering them by their semantic ``strength'', which we believe will reveal useful in further research. Finally, we map the consistency semantics to different practical systems and research prototypes. The scope of this paper is restricted to non-transactional semantics that apply to single storage object operations. 
We focus on non-transactional storage systems as they have become increasingly popular in recent years due to their simple implementations and good scalability. As such, our paper complements the existing surveys done in the context of transactional, database consistency semantics (see, e.g., \cite{Adya:99}), which we omit for space limitations.

\vspace{0.2cm}
This survey is organized as follows.
In Section~\ref{sec:model} we define our model of a distributed system and set up the framework for reasoning about different consistency semantics. 
In order to ensure the broadest coverage of our work, 
we model the distributed system as asynchronous, 
i.e., without predefined constraints on timing of computation and communication. Our framework, which we derive from the work of \citeN{Burckhardt:14}, captures the 
dynamic aspects of a distributed system, through 
\emph{histories} and \emph{abstract executions} of such systems.
We define an execution as a set of actions (i.e., \emph{operations}) invoked by some processes 
on the storage objects through their interface.
To analyze executions we adopt the notion of history, i.e., the set of operations of a given execution.
Leveraging the information attached to the history, we are able 
to properly capture the intrinsic complexity of executions.
Namely, we can group and relate operations according to their features 
(e.g., by the processes and objects they refer to, and by their timings),
or by the dynamic relationships established during executions (e.g., causality).
Additionally, abstract executions augment histories with orderings of operations that account 
for the resolution of write conflicts and their propagation within the storage system.

Section~\ref{sec:nontrans} brings the main contribution of our paper: a survey of more than 50 different consistency semantics proposed in the context of non-transactional distributed storage systems.\footnote{Note that, while this paper focuses on survey of consistency semantics proposed in the context of distributed storage, our approach maintains generality as our consistency definitions are applicable to other replicated data structures beyond distributed storage.} We define many of these models employing the framework specified in Section~\ref{sec:model}, i.e., using declarative compositions of logic predicates over graph entities.
In turn, these definitions 
enable us to establish the hierarchical partial order of consistency semantics according to their semantic strengths (given in Figure~\ref{fig:non-trans_graph} of Section~\ref{sec:nontrans}). For better readability, we also loosely classify consistency semantics into ten \emph{families}, which group them by their common traits.

We  discuss our work in the context of related consistency surveys in Section~\ref{sec:relwork} and conclude in Section~\ref{sec:conclusion}.  We further complement our survey with a summary of all consistency predicates defined in this work (Appendix~\ref{sec:predicates}). In addition, for all consistency models mentioned in this work,
we provide references to their original, primary definitions, as well as pointers to research papers that propose related implementations (Appendix~\ref{sec:reftable}).
Specifically, we reference implementations that appeared in recent proceedings of the most relevant venues. We believe that this is a useful contribution on its own, as it will allow distributed systems researchers and, in particular, students, to navigate more easily through the very large number of research papers that deal with different subtleties of consistency.

\section{System model}
\label{sec:model}

In this section, we specify the main notions behind the reasoning about consistency semantics 
carried out in the rest of this paper. 
We rely on the concurrent objects abstraction, as presented by \citeN{Lynch.Tuttle:89} and by \citeN{Herlihy.Wing:90}, 
for the definitions of fundamental ``static'' elements of the system, such as objects and processes. Moreover, to reason about dynamic behaviors of the system (i.e., executions), we build upon the mathematical 
framework laid out in \cite{Burckhardt:14}.

\subsection{Preliminaries}

\paragraph{Objects and Processes}
We consider a distributed system consisting of a finite set of \emph{processes}, 
modeled as I/O automata \cite{Lynch.Tuttle:89}, 
interacting with \emph{shared} (or \emph{concurrent}) \emph{objects}  through a 
fully connected asynchronous communication network.  
Unless stated otherwise, processes and shared objects (or, simply, objects) are \emph{correct}, i.e., they do not fail. Each process and object is identified by a unique identifier.
We define $\mathit{ProcessIds}$ as the set of all process identifiers and $\mathit{ObjectIds}$ as the set of all object identifiers. 

Additionally, each object has a unique \emph{object type}.
Depending on the type, the object can assume \emph{values} belonging to a defined domain denoted by $\mathit{Values},$\footnote{For readability, we adopt a notation in which a set $\mathit{Values}$ is implicitly parametrized by object type.} 
and it supports a set of primitive \emph{operation types} (i.e., $\mathit{OpTypes} = \{rd, wr, inc, \ldots \}$)
that provide the only means to manipulate that object.
For simplicity and without loss of generality, unless specified otherwise, 
in this work we further classify operations as either 
\emph{reads} ($rd$) or \emph{writes} ($wr$). 
Namely, we model as a write (or \emph{update}) any operation that modifies the value of the object, while, conversely, 
reads return to the caller the current value held by the object's replica without
causing any change to it.
We adopt the term \emph{object replicas}, or simply \emph{replicas}, to refer to the different copies of a
same named shared object maintained in the storage system for  
fault tolerance or performance enhancement.
Ideally, replicas of the same shared object should hold the same data at any time.
The coordination protocols among replicas are however determined by the implementation of the shared object.

\paragraph{Time} Unless specified otherwise, we assume an asynchronous computation and communication model, 
with no bounds on computation and communication latencies.
However, when describing certain consistency semantics, we will be using terms such as \emph{recency} or \emph{staleness}. 
Such terms relate to the concept of \emph{real time}, i.e., an ideal and global notion of time that we use to reason about histories a posteriori, 
although it is not accessible by processes during executions. 
We refer to the real time domain as $\mathit{Time}$, which we model as the set of positive real numbers, i.e., $\mathcal{R}^+$.

\subsection{Operations, Histories and Abstract Executions}

\paragraph{Operations}
We describe an operation issued by a process on a shared object as the tuple
$(proc, type, obj, ival, oval, stime, rtime)$, where:
\begin{itemize}
\item $proc \in \mathit{ProcessIds}$ is the id of the process invoking the operation.
\item $type \in \mathit{OpTypes}$ is the operation type.
\item $obj \in \mathit{ObjectIds}$ is the id of the object on which the operation is invoked.
\item $ival \in \mathit{Values}$ is the operation input value.
\item $oval \in \mathit{Values} \cup \{\nabla\}$ is the operation output value, or $\nabla$ if the operation does not  return.
\item $stime \in \mathit{Time}$ is the operation invocation time.
\item $rtime \in \mathit{Time} \cup \{\Omega\}$ is the operation return time, or $\Omega$ if the operation does not  return.
\end{itemize}
By convention, we use the special value $\sqcup \in \mathit{Values}$ to represent the input value (i.e., $ival$) of reads
and, possibly, the return value (i.e., $oval$) of writes.
For simplicity, given operation $op$, we will use the notation $op.par$ to access its parameter 
named $par$ as expressed in the tuple (e.g., $op.type$ represents its type, and $op.ival$ its input value).

\paragraph{Histories}
A \emph{history} $H$ is a set of operations. Intuitively, a history contains all operations invoked in a given execution. We further denote by $H|_{wr}$ (resp., $H|_{rd}$) the set of write (resp., read) operations 
of a given history $H$ (e.g., $H|_{wr}=\{op\in H: op.type=wr\}$).

We further define the following relations on elements of a history:\footnote{For better readability, we implicitly assume relations are parametrized by a history.}  
\begin{itemize}
\item $rb$ (\emph{returns-before}) is a natural partial order on $H$ based on real-time precedency.
Formally: $rb \triangleq \{ (a, b) :  a, b \in H \wedge a.rtime < b.stime \}$.
\item $ss$ (\emph{same-session}) is an equivalence relation on $H$ that groups pairs of operations invoked by the same process --- we say such operations belong to the same \emph{session}. Formally: $ss \triangleq \{ (a, b) :  a, b \in H \wedge a.proc = b.proc \}$.
\item  $so$ (\emph{session order}) is a partial order defined as: $so \triangleq rb \cap ss$.
\item $ob$ (\emph{same-object}) is an equivalence relation on $H$ that groups pairs of operations invoked on the same object.
Formally: $ob \triangleq \{ (a, b) :  a, b \in H \wedge a.obj = b.obj \}$.
\item  $\mathit{concur}$ as the symmetric binary relation designating all pairs of real-time \emph{concurrent} operations invoked on the same object. Formally: 
$
concur \triangleq ob \setminus rb
$.
\end{itemize} 
For $(a,b)\in rel$ we sometimes alternatively write $a \xrightarrow[]{\text{\textit{rel}}} b$. We further denote by $rel^{-1}$ the inverse relation of $rel$. For the sake of a more compact notation, we use binary relation projections.
For instance, $rel|_{wr \rightarrow rd}$ identifies all pairs of operations belonging to $rel$ 
consisting of a write and a read operation.
Furthermore, if $rel$ is an equivalence relation, we adopt the notation
$a \approx_{rel} b \triangleq [a \xrightarrow[]{\text{\textit{rel}}} b]$.
We recall that an equivalence relation $rel$ on set $H$ partitions $H$ into equivalence classes
$[a]_{rel} = \{b \in H : b \approx_{rel} a\}$. We write $H/\approx_{rel}$ to denote the set of equivalence classes determined by $rel$. 
We complement the $concur$ relation  with the function $\mathit{Concur}: H\rightarrow 2^H$ to denote the set of write operations concurrent with a given operation:
\begin{equation}
\mathit{Concur(a)} \triangleq \{b \in H|_{wr}: (a,b)\in \mathit{concur}\}
\end{equation}

\paragraph{Abstract executions} We model system executions using the concept of 
\emph{abstract execution}, following \citeN{Burckhardt:14}. An abstract execution is a multi-graph $\A = (H,vis,ar)$ built on a given history $H$, which it complements with two relations on elements of $H$, i.e., $vis$ and $ar$. Whereas histories describe the observable outcomes of executions,
$vis$ and $ar$, intuitively, capture the non-determinism of the asynchronous environment (e.g., message delivery order), as well as implementation-specific constraints (e.g., conflict-resolution policies).
In other words, $vis$ and $ar$ determine the relations between pairs of operations in a 
history that explain and justify its outcomes. 
More specifically:
\begin{itemize}
\item $vis$ (\emph{visibility}) is an acyclic natural relation that accounts for the propagation of write operations. 
Intuitively, $a$ be visible to $b$ (i.e., $a \xrightarrow[]{\text{\textit{vis}}} b$) means that the effects of $a$ are visible to the process invoking $b$ (e.g., $b$ may read a value written by $a$). 
Two write operations are \emph{invisible} to each other if they are not ordered by $vis$.

\item $ar$ (\emph{arbitration}) is a \emph{total} order on operations of the history that specifies 
how  the system resolves conflicts due to concurrent and invisible operations. In practice, such total order can be achieved in various ways: through the 
adoption of a distributed timestamping \cite{Lamport:78} or consensus protocol \cite{Birman.ea:91,Hadzilacos.Toueg:94,Lamport:01}, 
using a centralized serializer, or 
using a deterministic conflict resolution policy.
\end{itemize}

Depending on constraints expressed by $vis$,
during an execution processes may observe different orderings of write operations, which we call \emph{serializations}. 

We further define the \emph{happens-before} order ($hb$) as the transitive closure 
of the union of $so$ and $vis$, denoted by: 
\begin{equation}
\label{eq:hb}
hb \triangleq (so \cup vis)^+
\end{equation}

\subsection{Replicated data types and return value consistency}
Rather than defining the current system state as a set of values held by shared objects,
following \citeN{Burckhardt:14},
we employ a graph abstraction called (operation) \emph{context}, which
encodes the information 
of an abstract execution $\A$, taking a projection on visibility ($vis$) 
with respect to a given operation $op$.
Formally, given $\mathcal{C}$ as the set of contexts of all operations in a given abstract execution $\A$, we define the context of an operation $op$ as:
\begin{equation}
\label{eq:cxt}
C = cxt(\A,op) \triangleq \A|_{op,vis^{-1}(op),vis,ar}\end{equation}

Further, we adopt the concept of \emph{replicated data type} \cite{Burckhardt:14}
to define the type of shared object implemented in the distributed system
(e.g., read/write register, counter, set, queue, etc.).
For each replicated data type, a function $\mathcal{F}$ specifies 
the set of intended return values of an operation $op \in H$ in relation to 
its context, i.e.,  $\mathcal{F}(op, cxt(\A,op))$.
Using $\mathcal{F}$, we can define \emph{return value consistency} as:
\begin{equation} 
\RVAL \triangleq \forall op \in H : op.oval \in \mathcal{F}(op, cxt(\A,op))
\end{equation}
Essentially, return value consistency is a predicate on abstract executions that guarantees 
that the return value of any given operation of that execution will belong to the set of its intended return values.

Given operation $b\in H$ and its context $cxt(\A,b)$, 
we let $a = prec(b)$ be the (unique) latest operation preceding $b$
in $ar$, such that: $a.oval \neq \nabla \wedge a \in H|_{wr}\cap vis^{-1}(b)$. In other words, $prec(b)$ is the last write visible to $b$ according to the ordering specified by $ar$. 
If no such preceding operation exists (e.g., if $b$ is the first operation of the execution according to $ar$), 
by convention $prec(b).ival$ is a default value equal to $\bot$.

In this paper we adopt the read/write register (i.e., read/write storage) as reference replicated data type, 
which is defined by the following intended return value function:
\begin{equation}
\label{eq:Freg}
\mathcal{F}_{reg}(op, cxt(\A,op)) = prec(op).ival
\end{equation}

Note that, while the focus of this survey is on read/write storage, the consistency predicates defined in this paper take $\mathcal{F}$ as a parameter, and therefore directly extend to other replicated data types. 

\subsection{Consistency semantics}

Following \citeN{Burckhardt:14}, we define \emph{consistency semantics},  
sometimes also called \emph{consistency guarantees},  as conditions on attributes
and relations of abstract executions, expressed as first-order logic predicates.
We write $\A \models \mathcal{P}$ if the consistency predicate $\mathcal{P}$ is true for abstract execution $\A$. 
Hence, defining a consistency model amounts to collecting all the required consistency predicates 
and then specifying that histories must be justifiable by at least an abstract execution 
that satisfies them all.

Formally, given history $H$ and $\mathcal{\A}$ as the set of all possible abstract executions on $H$,
we say that history $H$ satisfies some consistency predicates 
$\mathcal{P}_1, \dotsc \mathcal{P}_n$
if it can be extended to some abstract execution that satisfies them all:
\begin{equation} 
H \models \mathcal{P}_1 \wedge \dotsb \wedge \mathcal{P}_n \Leftrightarrow \exists \A \in \mathcal{\A} : \mathcal{H}(\A) = H \wedge \A \models \mathcal{P}_1 \wedge \dotsb \wedge \mathcal{P}_n
\end{equation}
In the notation above, given the abstract execution $\A = (H,vis,ar)$, $\mathcal{H}(\A)$ denotes $H$.

 \section{Non-transactional consistency semantics}
\label{sec:nontrans}

\begin{figure}[!htp]
	\centering
	\includegraphics[angle=-90,width=0.65\textwidth]{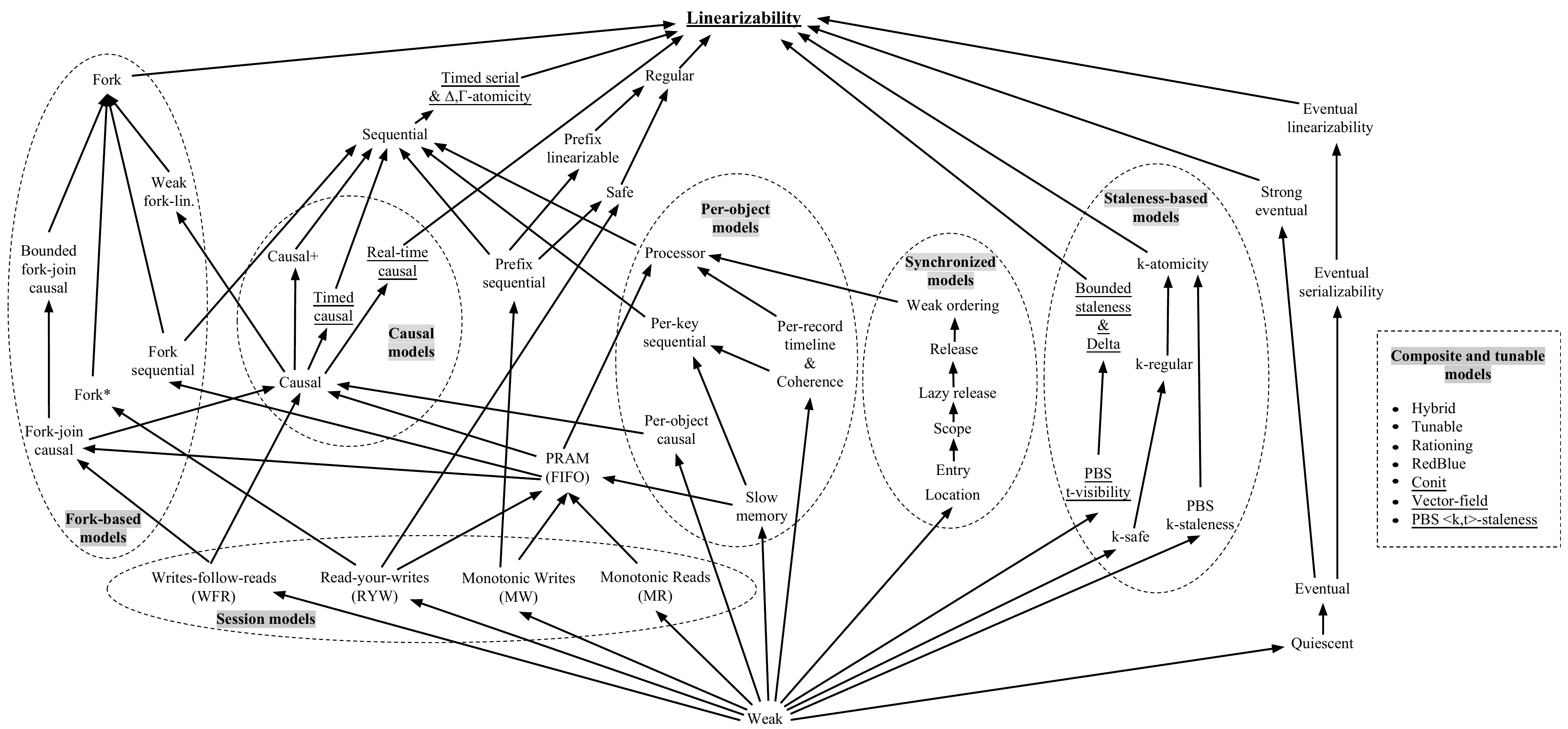} 	\caption{\footnotesize Hierarchy of non-transactional consistency models. \\A directed edge from  consistency semantics A to consistency semantics B means that any execution that satisfies B also satisfies A.
	Underlined models explicitly reason about timing guarantees.}
	\label{fig:non-trans_graph}
\end{figure}

In this section we analyze and survey the consistency semantics of systems which adopt single operations as their primary  operational constituent (i.e., non-transactional consistency semantics).
The consistency models described in the rest of the paper appear in Figure \ref{fig:non-trans_graph},
a comprehensive graph that proposes a partial ordering of consistency semantics according to their semantic strength, as well as a more loosely defined clustering into \emph{families} of consistency models. This classification draws both from strength of different consistency semantics  and from the underlying common factors that underpin their definitions.

In the remainder of this section we examine each family of consistency semantics.
Section~\ref{subsec:lin} introduces linearizability and other strong consistency models, while in Section~\ref{subsec:weak} we consider eventual and weak consistency.
Next we analyze PRAM and sequential consistency (Section~\ref{subsec:pram-seq}), and, in Section~\ref{subsec:session}, the models based on the concept of session.
Section~\ref{subsec:causal} proposes an overview of consistency semantics explicitly dealing with causality,
while in Section~\ref{subsec:timed} we study staleness-based models. This is followed by an overview of fork-based models (Section~\ref{subsec:fork}).
Section~\ref{subsec:tunable} and \ref{subsec:perobject} respectively deal with tunable and per-object semantics.
Finally, we survey the family of consistency models based on synchronization primitives (Section~\ref{subsec:sync}).

\subsection{Linearizability and related ``strong" consistency semantics}
\label{subsec:lin}
The gold standard and the central consistency model for non-transactional systems is \textbf{linearizability}, defined by \citeN{Herlihy.Wing:90}. Roughly speaking, linearizability is a correctness condition that 
establishes that each operation shall appear to be applied instantaneously 
at a certain point in time between its invocation and its response.
Linearizability, often informally dubbed \emph{strong consistency},\footnote{Note that the adjective ``strong'' 
	has also been used in literature to identify indistinctly 
	\emph{linearizability} and \emph{sequential consistency} 
	(which we define in Section \ref{subsec:pram-seq}), 
	as they both entail single-copy-semantics 
	and require that a single ordering of operations be observed by all processes.}
has been for long regarded as the ideal 
correctness condition at which distributed storage implementations should aim.
Linearizability features a
\emph{locality} property: a composition of linearizable objects is itself linearizable -- hence, linearizability enables modular design and verification. 

Although very intuitive to understand, the strong semantics of linearizability make it challenging to implement. 
In this regard, \citeN{Gilbert.Lynch:02}, formally proved the \emph{CAP} theorem, an assertion informally presented in previous works \cite{RFC0677,Davidson.Garcia-Molina.ea:85,Coan.ea:86,Brewer:00}, that binds linearizability to the ability of a system of maintaining a non-trivial level of availability when confronted with network partitions.
In a nutshell, the \emph{CAP} theorem states that in presence of network partitions a distributed storage system has to sacrifice either availability or linearizability.

\citeN{Burckhardt:14} breaks down linearizability into three components:
\begin{equation}
\textsc{Linearizability}(\mathcal{F}) \triangleq \textsc{SingleOrder} \wedge \textsc{RealTime} \wedge \RVAL \end{equation}
where:
\begin{equation}
\label{eq:singleorder}
\textsc{SingleOrder} \triangleq \exists H' \subseteq \{op \in H : op.oval = \nabla\}: vis = ar \setminus(H' \times H)
\end{equation}
and
\begin{equation}
\label{eq:realtime}
\textsc{RealTime} \triangleq rb \subseteq ar
\end{equation}
In other words, \textsc{SingleOrder} imposes a single global order that defines both $vis$ and $ar$,
whereas \textsc{RealTime} constrains arbitration ($ar$) to comply to the returns-before partial ordering ($rb$). Finally, $\RVAL$ specifies the return value consistency of a replicated data type. We recall that, as per Eq.~\ref{eq:Freg}, in case of read/write storage this is the value written by the last write (according to $ar$) visible to a given read operation $rd$.

\begin{figure}[h]
	\centering
	\includegraphics[width=0.60\textwidth]{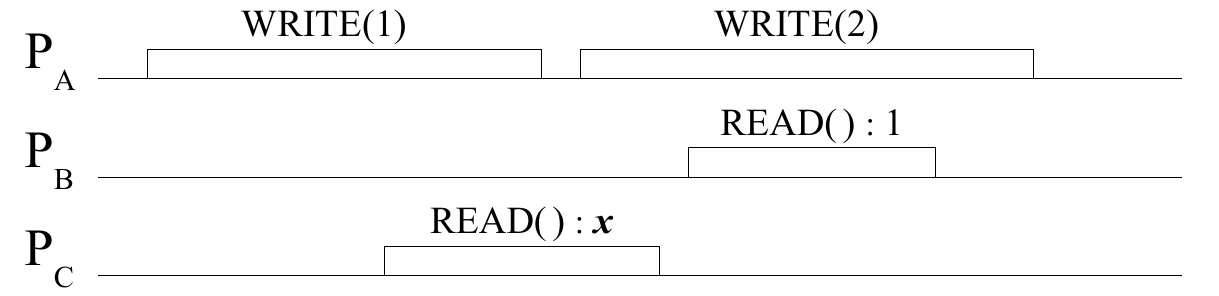} 	\caption{An execution exhibiting read-write concurrency (real time flows from left to right). The register is initialized to 0.  
	Atomic (linearizable) semantics allow \emph{x} to be 0 or 1. 
	Regular semantics allow \emph{x} to be 0, 1 or 2. With safe semantics \emph{x} may be any value.}
	\label{fig:reg-ex}
\end{figure}

A definition tightly related to that one of linearizability had been previously 
provided by \citeN{Lamport:86:vol2} for the \textbf{atomic} register semantic.
Lamport describes a single-writer multi-reader (SWMR) shared register to be atomic 
\emph{iff} each read operation not overlapping 
a write returns the last value actually written on the register, 
and the read values are the same as if the operations had been performed sequentially (i.e., without overlapping). Essentially, this definition implies the existence of a point in time 
(the \emph{linearization point}) at which each operation is actually applied on the shared register.\footnote{The existence of an instant 
at which each operation becomes atomically \emph{visible} had originally been postulated by \citeN{Lamport:83}.}
It is easy to show that 
atomicity and linearizability are equivalent for read-write registers.
However, linearizability is a more general condition designed 
for generic shared data structures that allow for a broader set of operational semantics 
than those offered by registers.
Besides atomic registers, \citeN{Lamport:86:vol2} defines  two slightly weaker semantics for SWMR registers: \textbf{safe} and \textbf{regular}.
In absence of read-write concurrency, they both guarantee that a read returns the last written value, 
exactly like the atomic semantics. The difference between the three resides in the allowed set of return values for a read operation 
concurrent with a write.
Namely, with a safe register, a read concurrent with some write may return any value. 
On the other hand, with a regular register, a read operation concurrent with some writes 
may return either the value written by the most recent complete write, 
or a value written by a concurrent write.
This difference is illustrated in Figure \ref{fig:reg-ex}.

Formally, regular and safe semantics can be defined as follows:
\begin{equation}
\label{eq:regular}
\textsc{Regular}(\mathcal{F}) \triangleq \textsc{SingleOrder} \wedge \textsc{RealTimeWrites} \wedge \RVAL \end{equation}
\begin{equation}
\label{eq:safe}
\textsc{Safe}(\mathcal{F}) \triangleq \textsc{SingleOrder} \wedge \textsc{RealTimeWrites} \wedge \textsc{Seq}\RVAL \end{equation}
where
\begin{equation}
\label{eq:realtimewrites}
\textsc{RealTimeWrites} \triangleq rb|_{wr \rightarrow op} \subseteq ar
\end{equation}
is a restriction of real-time ordering only for writes (preceding reads or other writes), and 
\begin{equation} 
\textsc{Seq}\RVAL \triangleq \forall op \in H : \mathit{Concur(op)}=\emptyset \Rightarrow op.oval \in \mathcal{F}(op, cxt(\A,op))
\end{equation}
which restricts the return value consistency only to read operations that are not concurrent with any write.

\subsection{Weak and eventual consistency}
\label{subsec:weak}
At the opposite end of the consistency spectrum lies \textbf{weak} consistency.
Although this term has been traditionally used in literature to identify any consistency model weaker than sequential consistency,
recent works \cite{Vogels:08,Bermbach.Kuhlenkamp:13} associate it 
to a more specific albeit rather vague definition: 
a weakly consistent system does not guarantee that reads return the most recent value written, 
and several (often underspecified) requirements have to be satisfied for a value to be returned.
In effect, weak consistency does not provide ordering guarantees -- 
hence, no synchronization protocol is actually required.
Even though this model might seem to have limited usability, 
it is in fact implemented in situations in which having a synchronization protocol would
be too costly, and a fortuitous exchange of information between
replicas can be good enough. 
For example, a typical use case for weak consistency are the relaxed caching policies that can be applied  across various tiers of a web application, or even the cache implemented in web browsers.

\textbf{Eventual} consistency is a slightly stronger notion than weak consistency. Namely, under eventual consistency, replicas converge towards identical copies in the absence of further updates.
In other words, 
if no new write operations are invoked on the object, \emph{eventually} all reads will return the same value.
Eventual consistency was first defined by \citeN{Terry.Demers.ea:94} and then further popularized more than a decade later by 
\citeN{Vogels:08} with the advent of
highly available storage systems (i.e., \emph{AP} systems in the CAP 
theorem parlance).
Eventual consistency is especially suited in contexts where coordination 
is not practical or too expensive (e.g., in mobile and wide area settings) \cite{Saito.Shapiro:05}. 
Despite its wide adoption, eventual consistency leaves to the application programmer the burden of dealing with transient \emph{anomalies} -- i.e., behaviors deviating from that of an ideal linearizable execution.
Hence, a quite large body of recent work has been aiming to achieve a better understanding of its subtle implications
\cite{Bermbach.Tai:11,Bernstein.Das:13,Bailis.Ghodsi:13,Bailis.ea:14}. At its core, eventual consistency constrains replicas' eventual state (i.e., their \emph{convergence}): 
in fact it does not provide any guarantees about recency and ordering of operations.
\citeN{Burckhardt:14} proposes a formal definition of eventual consistency:
\begin{multline}
\textsc{EventualConsistency}(\mathcal{F}) \triangleq \\ \textsc{EventualVisibility} \wedge \textsc{NoCircularCausality} \wedge \RVAL
\end{multline}
where:
\begin{multline}
\textsc{EventualVisibility} \triangleq \forall a \in H, \forall [f] \in H/\approx_{ss} : \\
|\{ b \in [f] : (a \xrightarrow[]{\text{\textit{rb}}} b) \wedge (a \xnrightarrow{vis} b)\}| < \infty 
\end{multline}
and
\begin{equation}
\textsc{NoCircularCausality} \triangleq \mathit{acyclic(hb)}
\end{equation}
that is, the acyclic projection of $hb$, defined in Eq. \ref{eq:hb}.
\textsc{EventualVisibility} mandates that, eventually, operation $op$ will be visible to another operation  $op'$ invoked after the completion of $op$.

In an alternative attempt at clarifying the definition of eventual consistency, 
\citeN{Shapiro.ea:11} identify the following properties from replicas' viewpoint:
\begin{itemize}
\item \emph{Eventual delivery}: if some correct replica applies a write operation $op$, $op$ is eventually applied by all correct replicas;
\item \emph{Convergence}:
all correct replicas that have applied the same write operations eventually reach equivalent state;
\item \emph{Termination}: 
all operations complete. \end{itemize}
To this definition of eventual consistency, \citeN{Shapiro.ea:11} add the following constraint:
\begin{itemize}
\item \emph{Strong convergence}:
all correct replicas that have applied the same write operations \emph{have} equivalent state.
\end{itemize}
In other words, this last property guarantees that any two replicas that have applied the same (possibly unordered) set of writes will hold the same data. A storage system enforcing both eventual consistency and strong convergence is said to implement \textbf{strong eventual} consistency.

We capture strong convergence from the perspective of read operations, by requiring that reads which have the identical sets of visible writes return the same values. 
\begin{multline}
\label{eq:strongcon}
\textsc{StrongConvergence} \triangleq \forall a,b\in H|_{rd}: vis^{-1}(a)|_{wr}=vis^{-1}(b)|_{wr} \Rightarrow a.oval = b.oval
\end{multline}
Then, strong eventual consistency can be defined as:
\begin{multline}
\textsc{StrongEventualConsistency}(\mathcal{F}) \triangleq \\ \textsc{EventualConsistency}(\mathcal{F}) \wedge \textsc{StrongConvergence} 
\end{multline}

\textbf{Quiescent} consistency \cite{Herlihy.Shavit:08} requires that 
if an object stops receiving updates (i.e., becomes quiescent), 
then the execution is equivalent to some sequential execution containing only complete operations.
Although this definition resembles eventual consistency, 
it does not guarantee termination: 
a system that does not stop receiving updates will not reach quiescence, thus replicas convergence.
Following \cite{Burckhardt:14}, we formally define quiescent consistency as:
\begin{multline}
\textsc{QuiescentConsistency}(\mathcal{F}) \triangleq 
|H|_{wr}| < \infty \Rightarrow \\
\exists C \in \mathcal{C} : \forall[f] \in H/\approx_{ss} : |\{op \in [f]: op.oval \notin \mathcal{F}(op,C) \}| < \infty
\end{multline}

\subsection{PRAM and sequential consistency}
\label{subsec:pram-seq}
Pipeline RAM (\textbf{PRAM} or FIFO) consistency  \cite{Lipton.Sandberg:88} prescribes that
all processes see write operations issued by a given process in the same order as they were invoked by that process.
On the other hand, processes may observe writes issued by different processes in different orders. 
Thus, no global total ordering is required. 
However, the writes from any given process (session) must be serialized in order, 
as if they were in a pipeline -- hence the name. 
We define PRAM consistency by requiring the visibility partial order to be a superset of session order: 
\begin{equation}
\textsc{PRAM} \triangleq so \subseteq vis
\end{equation}
As proved by \citeN{Brzezinski.Sobaniec.ea:03}, PRAM consistency is ensured \emph{iff} 
the system provides read-your-write, monotonic reads and monotonic writes guarantees,
which we will introduce in Section \ref{subsec:session}.

In a storage system implementing \textbf{sequential} consistency all operations are serialized 
in the same order on all replicas, and the ordering of operations determined by each process is preserved.
Formally:
\begin{equation}
\textsc{SequentialConsistency}(\mathcal{F}) \triangleq \textsc{SingleOrder} \wedge \textsc{PRAM} \wedge \RVAL
\end{equation}
Thus, sequential consistency, first defined in \cite{Lamport:79}, is a guarantee of 
ordering rather than recency. Like linearizability, sequential consistency enforces a common global order of operations.
Unlike linearizability, sequential consistency 
does not require real-time ordering of operations across different sessions: only the real-time 
ordering of operations invoked by the same process is preserved 
(as in PRAM consistency).\footnote{In Section \ref{subsec:sync} we present \emph{processor} consistency: 
a model whose semantic strength stands between those of PRAM and sequential consistency.}
A quantitative comparison of the power and costs involved 
in the implementation of sequential consistency and linearizability is presented by \citeN{Attiya:Welch:94}.

\begin{figure}[h]
	\centering
	\includegraphics[width=0.60\textwidth]{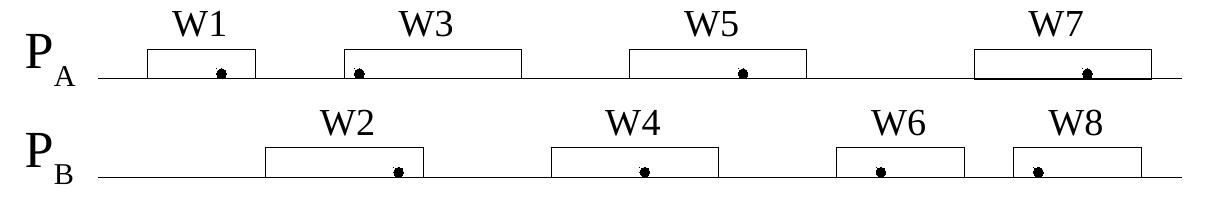} 	\caption{An execution with processes issuing write operations on a shared object.\\
	Black spots are the chosen linearization points.}
	\label{fig:pram_seq-ex}
\end{figure}
Figure \ref{fig:pram_seq-ex} shows an execution featuring two processes issuing 
write operations on a shared object.
Let us suppose that the two processes also continuously perform read operations. Each process will observe a certain serialization of the write operations.
If we were to assume that the system respects PRAM consistency, those two processes might observe, for instance, 
the following two serializations:
\begin{align} 
\label{eq:seqprama}
S_{P_{A}} : \quad W1 \quad W2 \quad W3 \quad W5 \quad W4 \quad W7 \quad W6 \quad W8\tag{S.1}\\
\label{eq:seqpramb}
S_{P_{B}} : \quad W1 \quad W3 \quad W5 \quad W7 \quad W2 \quad W4 \quad W6 \quad W8\tag{S.2}
\end{align}
If the system implemented sequential consistency, then $S_{P_{A}}$ would be equal to $S_{P_{B}}$ and 
it would respect the ordering of operations imposed by each writing process.
Thus, any of (\ref{eq:seqprama}) or (\ref{eq:seqpramb}) would be acceptable.
On the other hand, assuming the system implements linearizability, and assigning linearization points as indicated by the points  
in Figure \ref{fig:pram_seq-ex}, (\ref{eq:srlin}) would be the only allowed serialization:
\begin{align} 
\label{eq:srlin}
S_{Lin} : \quad W1 \quad W3 \quad W2 \quad W4 \quad W5 \quad W6 \quad W8 \quad W7 \tag{S.3} 
\end{align}

\subsection{Session guarantees} 
\label{subsec:session}
\emph{Session guarantees} were first described by \citeN{Terry.Demers.ea:94}.
Although originally defined in connection to \emph{client} sessions, session guarantees may as well apply to situations in which the concept of session
is more loosely defined and it just refers to a specific process' point of view on the execution. 
We note that previous works in literature have classified session guarantees as \emph{client-centric models} \cite{Tanenbaum.Steen:07}. 

\textbf{Monotonic reads} states that successive reads must reflect a non-decreasing set of writes. 
Namely, if a process has read a certain value $v$ from an object, any successive read operation will not return any value written before $v$.
Intuitively, a read operation can be served only by those replicas that have executed all write operations whose effects have already been observed by the requesting process.
In effect, we can represent this by saying that, given three operations $a,b,c \in H$, if $a \xrightarrow[]{\text{vis}} b$
and $b \xrightarrow[]{\text{so}} c$, where $b$ and $c$ are read operations, then $a \xrightarrow[]{\text{vis}} c$, i.e., the transitive
closure of $vis$ and $so$ is included in $vis$.
\begin{multline}
\textsc{MonotonicReads} \triangleq  
\forall a\in H, \forall b,c \in H|_{rd} : a \xrightarrow[]{\text{vis}} b \wedge b \xrightarrow[]{\text{so}} c \Rightarrow a \xrightarrow[]{\text{vis}} c \\
\triangleq (vis;so|_{rd \rightarrow rd}) \subseteq vis
\end{multline}

\textbf{Read-your-writes} guarantee (also called read-my-writes \cite{Terry.Prabhakaran.ea:13,Burckhardt:14})
requires that a read operation invoked by a process can only be carried out by replicas 
that have already applied all writes previously invoked by the same process.
\begin{multline}
\textsc{ReadYourWrites} \triangleq 
\forall a\in H|_{wr}, \forall b \in H|_{rd} : a \xrightarrow[]{\text{so}} b \Rightarrow
a \xrightarrow[]{\text{vis}} b\\
\triangleq so|_{wr \rightarrow rd} \subseteq vis
\end{multline}

\begin{figure}[h]
	\centering
	\includegraphics[width=0.60\textwidth]{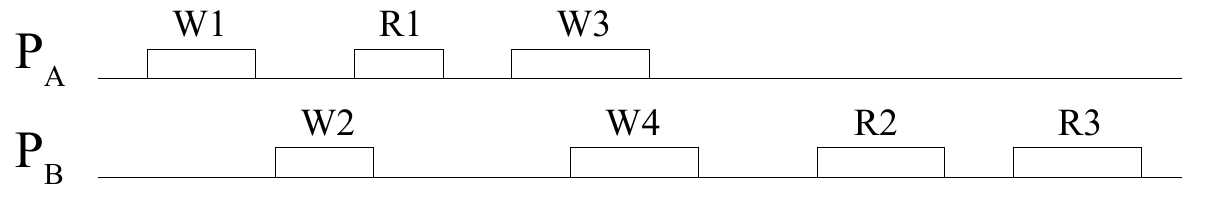} 	\caption{An execution with processes issuing read and write operations on a shared object.}
	\label{fig:pram_ryw-ex}
\end{figure}

Let us assume that two processes issue read and write operations on a shared object as in Figure \ref{fig:pram_ryw-ex}.

\noindent Given such execution, $P_A$ and $P_B$ could observe the following serializations, 
which satisfy the read-your-write guarantee but not PRAM consistency:

\begin{align} 
S_{P_{A}} : \quad W1 \quad W3 \quad W4 \quad W2 \tag{S.4}\\ 
S_{P_{B}} : \quad W2 \quad W4 \quad W3 \quad W1 \tag{S.5}
\end{align}

We note that some works in literature refer to \emph{session consistency} as a special case of
read-your-writes consistency that can be attained through \emph{sticky} client sessions, i.e., those sessions in which the process always invokes operations on a given replica.

In a system that ensures \textbf{monotonic writes} a write is only performed on a replica if the replica has already performed all previous writes of the same session. 
In other words, replicas shall apply all writes belonging to the same session according to the order in which they were issued.
\begin{multline}
\textsc{MonotonicWrites} \triangleq 
\forall a, b \in H|_{wr} : a \xrightarrow[]{\text{so}} b 
\Rightarrow a \xrightarrow[]{\text{ar}} b 
\triangleq so|_{wr \rightarrow wr} \subseteq ar
\label{eq:mw}
\end{multline}

\textbf{Writes-follow-reads}, sometimes called \emph{session causality}, 
is somewhat the converse concept of read-your-write guarantee as it ensures that writes
made during the session are ordered after any writes made 
by any process on any object whose effects were seen by previous reads in the same session.
\begin{multline}
\textsc{WritesFollowReads} \triangleq 
\forall a,c\in H|_{wr}, \forall b\in H|_{rd} : a \xrightarrow[]{\text{vis}} b \wedge b \xrightarrow[]{\text{so}} c \Rightarrow a \xrightarrow[]{\text{ar}} c \\
\triangleq (vis;so|_{rd \rightarrow wr}) \subseteq ar
\end{multline}

We note that some of the session guarantees embed specific notions of causality, 
and that in fact, as proved by \citeN{Brzezinski.ea:04},
causal consistency -- which we describe next -- requires and includes them all.

\subsection{Causal models}
\label{subsec:causal}
The commonly accepted notion of potential causality in distributed systems 
has been enclosed in the definition of 
the \emph{happened-before} relation introduced by \citeN{Lamport:78}.
According to this relation, two operations $a$ and $b$ are ordered if 
(a) they are both part of the same thread of execution, (b) $b$ reads a value written by $a$, or
(c) they are related by a transitive closure leveraging (a) and/or (b).
This notion, originally defined in the context of message passing systems, 
has been translated to a consistency condition 
for shared-memory systems by \citeN{Hutto.Ahamad:90}.
The potential causality relation establishes a partial order over operations
which we represent as $hb$ in (\ref{eq:hb}).
Hence, while operations that are potentially causally\footnote{While the most appropriate terminology would 
be ``potential causality'', for simplicity, hereafter we will use ``causality''.}
related must be seen by all processes in the same order, operations that are 
not causally related (i.e., causally concurrent) may be observed in different orders by different processes. 
In other words, \textbf{causal} consistency  
dictates that all replicas agree on the ordering of causally related operations \cite{Hutto.Ahamad:90,Ahamad.Neiger.ea:95,P-Mahajan.Dahlin:11}.
This can be expressed as the conjunction of two predicates \cite{Burckhardt:14}:
\begin{itemize}
\item $\textsc{CausalVisibility} \triangleq hb \subseteq vis$
\item $\textsc{CausalArbitration} \triangleq hb \subseteq ar$
\end{itemize}
Hence, causal consistency is defined as:
\begin{multline}
\textsc{Causality}(\mathcal{F}) \triangleq \textsc{CausalVisibility} \wedge \textsc{CausalArbitration} \wedge \RVAL
\end{multline}

\begin{figure}[h]
	\centering
	\includegraphics[width=0.60\textwidth]{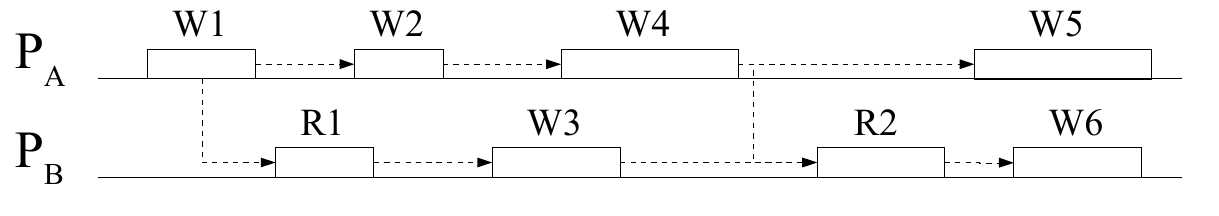} 	\caption{An execution with processes issuing operations on a shared object.\\
	Arrows express causal relationships between operations.}
	\label{fig:pram_causal-ex}
\end{figure}

Figure \ref{fig:pram_causal-ex} represents an execution with two processes writing and reading the value
of a shared object, with arrows indicating the causal relationships between operations.

\noindent Assuming the execution respects PRAM but not causal consistency, we might have the following serializations:
\begin{align} 
S_{P_{A}} : \quad W1 \quad W2 \quad W4 \quad W5 \quad W3 \quad W6 \tag{S.6}\\ 
S_{P_{B}} : \quad W3 \quad W6 \quad W1 \quad W2 \quad W4 \quad W5 \tag{S.7}
\end{align}
Otherwise, with causal consistency (which implies PRAM), we could have obtained these serializations:
\begin{align} 
S_{P_{A}} : \quad W1 \quad W3 \quad W2 \quad W4 \quad W5 \quad W6 \tag{S.8}\\ 
S_{P_{B}} : \quad W1 \quad W2 \quad W3 \quad W4 \quad W6 \quad W5 \tag{S.9}
\end{align}

Recent work by \citeN{Bailis.Fekete.ea:12} promotes the use of explicit application-level causality, 
which is a subset of potential causality,\footnote{As argued in \cite{Bailis.Fekete.ea:12},
the application-level causality graph would be smaller in fanout and depth with respect to 
the traditional causal one, because it would only enclose relevant causal relationships, 
hinging on application-level knowledge and user-facing outcomes.}
for building highly available distributed systems that would entail less overhead
in terms of coordination and metadata maintenance.
Furthermore, an increasing body of research has been drawing attention on causal consistency, considered
as an optimal tradeoff between user-perceived correctness and coordination overhead,
especially in mobile or geo-replicated applications \cite{Lloyd.Freedman.ea:11,Bailis.Ghodsi.ea:13,Zawirski.ea:15}. 

\textbf{Causal+} (or convergent causal) consistency \cite{Lloyd.Freedman.ea:11} mandates, 
in addition to causal consistency, 
that all replicas should eventually and independently agree on conflicts resolution.
In fact, causally concurrent write operations may generate conflicting outcomes
which in convergent causal consistent systems are handled in the same way by commutative and associative functions.
Essentially, causal+ strengthens causal consistency with strong convergence (see Equation~(\ref{eq:strongcon})), which mandates that all correct replicas that have applied the same write operations have equivalent state. In a sense, causal+ consistency augments causal consistency with strong convergence, in the vein strong eventual consistency \cite{Shapiro.ea:11} strengthens eventual consistency. Hence, causal+ consistency can be expressed as:

\begin{equation}
\textsc{Causal+}(\mathcal{F}) \triangleq \textsc{Causality}(\mathcal{F}) \wedge \textsc{StrongConvergence}
\end{equation}

\textbf{Real-time causal} consistency has been defined in \cite{P-Mahajan.Dahlin:11} 
as a stricter condition than causal consistency that
enforces an additional condition: causally concurrent write operations that do not 
overlap in real-time must be applied according to their real-time order.
\begin{equation}
\textsc{RealTimeCausality}(\mathcal{F}) \triangleq 
\textsc{Causality}(\mathcal{F}) \wedge \textsc{RealTime}
\end{equation}
where \textsc{RealTime} is defined as in (\ref{eq:realtime}).

We note that although \cite{Lloyd.Freedman.ea:11} classifies real-time causal consistency as stronger than causal+ consistency, they are actually incomparable, as real-time causality --- as defined in \cite{P-Mahajan.Dahlin:11} --- does not imply strong convergence. Of course, one can devise a variant of real-time causality that respects strong convergence as well. 

\citeN{Attiya.ea:15} define \emph{observable causal} consistency as a strengthening of causal consistency
for multi-value registers (MVR)
that enforces the exposure of concurrency between operations when this concurrency may be inferred by processes from their observations.
Observable causal consistency has also been proved to be the strongest consistency model satisfiable 
for a certain class of highly-available data stores implementing MVRs.

\subsection{Staleness-based models}
\label{subsec:timed}
Intuitively, staleness based models allow reads to return old, \emph{stale} written values. They provide stronger guarantees than eventually consistent semantics, but weak enough to allow for more efficient implementations than linearizability. In literature, two common metrics are employed to measure staleness: (real) time and data (object) versions.

To the best of our knowledge, the first formalization of a consistency model explicitly dealing with time-based staleness is proposed by \citeN{Singla.ea:97}
as \textbf{delta} consistency.
According to delta consistency, writes are guaranteed to become visible at most after $t+delta$ time units.
Moreover, delta consistency is defined in conjunction with an ordering criterion (which is reminiscent of the \emph{slow memory} consistency model, that we postpone to Section \ref{subsec:perobject}):
writes to a given object by the same process are observed in the same order by all processes, 
but no global ordering is enforced for writes to a given object by different processes.

In an analogous way, \emph{timed consistency} models, as defined by \citeN{Torres-Rojas.Ahamad.ea:99}, 
restrict the sets of values that read operations may return 
by the amount of time elapsed since the preceding writes.
Specifically, in a \emph{timed serialization} 
all reads occur \emph{on time}, i.e., they do not return stale values when 
there are more recent ones that have been available 
for more than $\Delta$ units of time -- $\Delta$ being a parameter of the execution.
In other words, similarly to delta consistency, if a write operation is performed at time $t$,
the value written by this operation must be visible by all processes by time $t + \Delta$. 

\citeN{Mahajan.Setty.ea:10} define a consistency condition named \textbf{bounded staleness} 
which at its core is very similar to that of timed and delta semantics: 
a write operation of a given process becomes visible to other processes no later than a fixed amount of time.
However, this definition is also related to the use of a periodic message (i.e., a \emph{beacon})
which allows each process to keep up with updates from other processes or \emph{suspect} of missing updates. 
The differences among delta consistency, timed reads and bounded staleness are in fact 
matter of subtle operational details that derive from the diverse contexts 
and practical purposes for which those models were developed.
Hence, we can describe in formal terms the core semantics expressed by delta consistency, timed consistency models and bounded staleness as the following condition:
\begin{multline}
\textsc{TimedVisibility}(\Delta) \triangleq  \forall a\in H|_{wr},\forall b \in H, \forall t \in \mathit{Time} : \\
a.rtime = t \wedge b.stime = t + \Delta \Rightarrow a \xrightarrow[]{\text{vis}} b
\end{multline}

\textbf{Timed causal} consistency \cite{Torres-Rojas.Meneses.05} 
guarantees that each execution respects the partial ordering of causal consistency 
and that all reads are \emph{on time}, with tolerance $\Delta$: 
\begin{align}
\textsc{TimedCausality}(\mathcal{F},\Delta) \triangleq 
\textsc{Causality}(\mathcal{F}) \wedge \textsc{TimedVisibility}(\Delta) \end{align}
As depicted in Figure \ref{fig:non-trans_graph}, due to the timed visibility term, timed causal is a semantic condition stronger than causal consistency. 
Similarly, \textbf{timed serial} consistency \cite{Torres-Rojas.Meneses.05} 
combines the real-time global ordering guarantee with the timed serialization constraint.
Hence, a timed serial consistent execution with $\Delta = 0$ would in fact be linearizable.

\noindent \citeN{Golab.Li.ea:11} describe \textbf{$\Delta$-atomicity}, a semantic condition which is in fact equivalent to timed serial consistency.
Namely, according to $\Delta$-atomicity read operations may return either the value written by the last preceding write, or the value of a write operation returned up to $\Delta$ time units ago. 
In a follow-up work \cite{Golab.ea:14}, the same authors propose a novel metric called $\Gamma$ which entails fewer assumptions and is more robust than $\Delta$ against clock skews.
The corresponding consistency semantics, \textbf{$\Gamma$-atomicity}, expresses, as $\Delta$-atomicity,
a ``deviation'' in time of a given execution from a linearizable one having the same operations' outcomes.

\begin{figure}[h]
	\centering
	\includegraphics[width=0.60\textwidth]{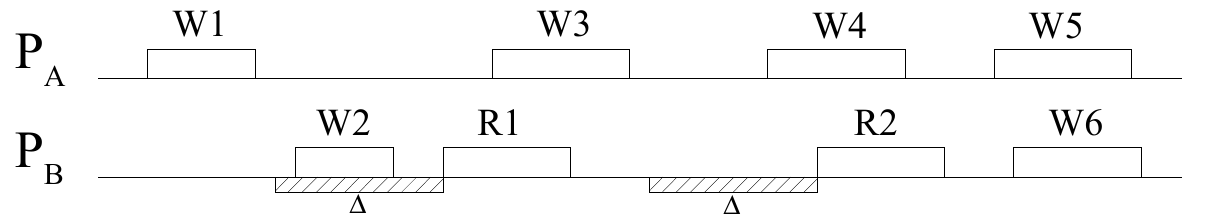} 	\caption{An execution with processes issuing operations on a shared object.\\
	Hatched rectangles highlight the $\Delta$ parameter of staleness-based read operations.}
	\label{fig:seq_timed-ex}
\end{figure}

\noindent We express the core notion of  $\Delta$-atomicity, $\Gamma$-atomicity and timed serial consistency in the following
predicate:
\begin{multline}
\textsc{TimedLinearizability}(\mathcal{F}, \Delta) \triangleq 
\\ \textsc{SingleOrder} \wedge \textsc{TimedVisibility}(\Delta) \wedge \RVAL
\end{multline}
Figure \ref{fig:seq_timed-ex} illustrates an execution featuring read operations of which outcomes should 
depend on a fixed timing parameter $\Delta$.

\noindent If we were to assume that, despite the timing parameter, $P_A$ and $P_B$ observed the following serialization:
\begin{align} 
S_{P_{A,B}} : \quad W2 \quad W6 \quad W1 \quad W3 \quad W4 \quad W5 \tag{S.10}
\end{align}
then such execution would be sequentially consistent but it would not satisfy timed serial consistency requirements.
Thus, this execution serves as hint of the relative strenghts of sequential and timed serial consistency models as
represented in Fig. \ref{fig:non-trans_graph}.

\textbf{Prefix} consistency \cite{Terry.ea:95,Terry:13}, 
also dubbed \textbf{timeline} consistency \cite{Cooper.Ramakrishnan.ea:08}, 
grants readers the guarantee of observing an ordered sequence of writes which nonetheless 
may not contain the most recent ones.
So it expresses a constraint in matter of ordering rather than recency of writes: 
the read value is the result of a specific sequence of writes upon whose order all replicas have agreed.
This pre-established order is supposedly reminiscent of that one imposed by sequential consistency. 
Thus, we could rename prefix consistency as \emph{prefix sequential} consistency,
whereas a version abiding real-time constraints would be called \emph{prefix linearizable} consistency. 
Formally, we describe prefix sequential consistency as:
\begin{align}
\textsc{PrefixSequential}(\mathcal{F}) \triangleq \textsc{SingleOrder} \wedge \textsc{MonotonicWrites} \wedge \RVAL
\end{align}
where the term named \textsc{MonotonicWrites} implies that the ordering of writes belonging to the same session is respected, as defined in (\ref{eq:mw}).
Similarly, we express prefix linearizable consistency as:
\begin{align}
\textsc{PrefixLinearizable}(\mathcal{F}) \triangleq \textsc{SingleOrder} \wedge \textsc{RealTimeWW} \wedge \RVAL
\end{align}
where
\begin{align}
\textsc{RealTimeWW} \triangleq rb|_{wr \rightarrow wr} \subseteq ar
\end{align}

In a study on quorum-based replicated systems with malicious faults, \citeN{Aiyer.ea:05} formalize relaxed semantics 
that tolerate limited version-based staleness.
Substantially, \textbf{$K$-safe}, \textbf{$K$-regular} and \textbf{$K$-atomic} (or \textbf{$K$-linearizability}) generalize 
the register consistency conditions previously introduced in \cite{Lamport:86:vol1}
and described in Section \ref{subsec:lin}, by permitting reads non-overlapping concurrent writes to return one of the latest $K$ values written. For instance \textbf{$K$-linearizability} can be formalized as:\footnote{Strictly speaking, $K$-linearizability implicitly assumes $K$ initial writes (i.e., writes with input value $\bot$) \cite{Aiyer.ea:05}.} 
\begin{multline}
\textsc{K-Linearizable}(\mathcal{F}, K) \triangleq \\ 
\textsc{SingleOrder} \wedge \textsc{RealTimeWW} \wedge \textsc{K-RealTimeReads}(K) \wedge \RVAL
\end{multline}
where
\begin{multline}
\label{eq:RTKreads}
\textsc{K-RealTimeReads}(K) \triangleq 
\forall a \in H|_{wr}, \forall b\in H|_{rd}, \forall PW \subseteq H|_{wr}, \forall pw\in PW: \\ 
|PW|<K \wedge a\xrightarrow[]{\text{\textit{ar}}}pw \wedge
pw\xrightarrow[]{\text{\textit{rb}}}b \wedge a\xrightarrow[]{\text{\textit{rb}}}b
\Rightarrow a\xrightarrow[]{\text{\textit{ar}}}b
\end{multline}

Finally, \citeN{Bailis.Venkataraman.ea:12} build on these results a series of
probabilistic models to predict the staleness of reads performed on eventually consistent quorum-based stores. They provide definitions of Probabilistically Bounded Staleness (PBS) \textbf{\emph{k}-staleness} and PBS \textbf{\emph{t}-visibility}. 
While the first describes a probabilistic model which restricts the staleness of values returned by read operations, the latter limits probabilistically the time before a write becomes visible. The combination of these two models is named PBS \textbf{$\langle$\emph{k, t}$\rangle$-staleness}.
In a sense, PBS k-staleness is a probabilistic weakening of $K$-atomicity, i.e., the one that with probability equal to 1 becomes $K$-linearizability. Similarly, PBS t-visibility is a probabilistic weakening of timed visibility.

\subsection{Fork-based models}
\label{subsec:fork}

Inherent trust limitations that arise in the context of outsourced storage and computations \cite{Cachin.Keidar.ea:09,Vukolic:10} has revamped the  research on algorithms and protocols expressly  conceived to deal with Byzantine faults \cite{Lamport.ea:82}, 
i.e faults that encompass arbitrary and malicious behavior. In the Byzantine fault model, 
faulty processes and shared objects may tamper data (within the limits of cryptography) or perform other arbitrary 
operations in order to deliberately disrupt executions.

Together with these algorithms, new consistency models were defined that reshaped the 
correctness conditions in accordance to what is actually attainable when coping 
with such strong fault assumptions. Whereas in the context of several untrusted storage repositories Byzantine fault tolerance could be applied to mask certain fault patterns \cite{Vukolic:10,Bessani.Correia.ea:13} and even implement strong consistency semantics (e.g., linearizability) \cite{Bessani.Mendes.ea:14,Dobre.ea:14}, when dealing with a single untrusted storage repository, the situation is different and the consistency needs to be relaxed \cite{Cachin.Keidar.ea:09}. Feasible consistency semantics in the context of interactions of correct clients with untrusted (Byzantine) storage have been captured within the family of \emph{fork-based} consistency models. In a nutshell, systems dealing with untrusted storage aim at providing linearizability when the storage is correct, but (gracefully) degrade to weaker consistency models, specifically, fork-based consistency models, when the storage exhibits a Byzantine fault.

The forefather of this family of models is \textbf{fork} (or fork-linearizable) consistency,
introduced by \citeN{Mazieres.Shasha:02}.
In short, a fork-linearizable system guarantees that if the storage system causes the visible histories of two
processes to differ even for just a single operation, they may never again observe each other's 
writes after that without the server being exposed as faulty.
Specifically, any divergence in the histories observed by different groups of correct processes
can be easily spotted by using any available communication protocol between them
(e.g., out-of-band communication, gossip protocols, etc.). Fork-linearizability respects session order (\textsc{PRAM} semantics) and real-time arbitration, thus can be expressed as follows:
\begin{align}
\textsc{ForkLinearizability}(\mathcal{F}) \triangleq \textsc{PRAM} \wedge \textsc{RealTime} \wedge \textsc{NoJoin} \wedge{\RVAL} 
\end{align}
\noindent where the \textsc{NoJoin} predicate stipulates that clients whose sequences of visible operations (also called \emph{views}) have been forked by an adversary, cannot be joined again:
\begin{multline}
\textsc{NoJoin} \triangleq \forall a_i,b_i,a_j,b_j\in H: 
a_i \not\approx_{ss} a_j \wedge 
(a_i,a_j)\in ar\setminus vis \wedge a_i \preceq_{so} b_i \wedge a_j \preceq_{so} b_j  \\ \Rightarrow 
(b_i,b_j),(b_j,b_i)\notin vis
\end{multline}

A subsequent model named \textbf{fork*} consistency was defined in \cite{Li.Mazieres:07} 
in order to allow the design of protocols that would offer better performance and liveness guarantees.
Fork* consistency relaxes the conditions of fork consistency 
by allowing forked groups of processes to observe at most one common 
operation issued by a certain correct process.
\begin{align}
\textsc{Fork*}(\mathcal{F}) \triangleq \textsc{ReadYourWrites} \wedge \textsc{RealTime} \wedge \textsc{AtMostOneJoin} \wedge{\RVAL} 
\end{align}
\noindent where
\begin{align}
\begin{aligned}
\textsc{AtMostOneJoin} \triangleq & \forall a_i,a_j\in H: 
a_i \not\approx_{ss} a_j \wedge  (a_i,a_j)\in ar\setminus vis \Rightarrow\\ 
 & \wedge |\{ b_i \in H: a_i \preceq_{so} b_i \wedge 
(\exists b_j\in H: a_j \preceq_{so} b_j \wedge b_i\xrightarrow[]{\text{\textit{vis}}}b_j\}| \le 1 \\
 & \wedge |\{ b_j \in H: a_j \preceq_{so} b_j \wedge 
(\exists b_i\in H: a_i \preceq_{so} b_i \wedge b_j\xrightarrow[]{\text{\textit{vis}}}b_i\}| \le 1 
\end{aligned}
\end{align}
Notice that, unlike fork-linearizability, fork* does not respect monotonicity of reads (and hence \textsc{PRAM}) \cite{Cachin.Keidar.ea:11}.

\textbf{Fork-sequential} consistency \cite{Oprea.Reiter:06,Cachin.Keidar.ea:09a} requires that 
whenever an operation becomes visible to multiple processes, 
all these processes share the same history of operations occurring before that operation.
Therefore, whenever a process reads a certain value written by another process,
the reader is guaranteed to share with the writer process the set of visible operation that precede that write operation.
Essentially, similarly to sequential consistency, a global order of operations is ensured
up to a common visible operation. Formally:
\begin{align}
\textsc{ForkSequential}(\mathcal{F}) \triangleq \textsc{PRAM} \wedge \textsc{NoJoin} \wedge{\RVAL} 
\end{align}

Mahajan et al. define \textbf{fork-join causal} consistency (FJC) as a
weaker variant of causal consistency that can preserve safeness and availability in spite of Byzantine faults \cite{Mahajan.Setty.ea:10}.
In a fork-join causal consistent storage system if a write operation $op$ issued by a correct process 
depends on a write operation $op'$ issued by any process, then, at every correct process, $op'$ becomes visible before $op$.
In other words, FJC enforces causal consistency among correct processes.
Besides, partitioned groups of processes are allowed to reconcile their histories through merging policies,
since inconsistent writes by a Byzantine process are treated as 
concurrent writes by multiple \emph{virtual processes}.
\textbf{Bounded fork-join causal} \cite{P-Mahajan.Dahlin:11} refines this clause by limiting
the number of forks accepted from a faulty node and thus bounding the number of virtual nodes
needed to represent each faulty node.

Finally, \textbf{weak fork-linearizability} \cite{Cachin.Keidar.ea:11} relaxes fork-linearizability conditions
in two ways: (1) after being partitioned in different groups, two processes may 
share the visibility of one more operation
(i.e., \emph{at-most-one-join}, as in fork* consistency) 
and (2) the real-time order of the last visible operation by each process may not be preserved (i.e., \emph{weak real-time order}).
These two conditions enable the design of protocols that allow for improved liveness guarantees (i.e., \emph{wait freedom}).
Weak fork-linearizability can be expressed as:
\begin{equation}
\textsc{WeakForkLin}(\mathcal{F}) \triangleq  
\textsc{PRAM} \wedge \textsc{K-RealTime(2)} \wedge \textsc{AtMostOneJoin} \wedge{\RVAL} 
\end{equation}
where \textsc{K-RealTime}(2) predicate is equivalent \textsc{K-RealTimeReads}(2) defined in Equation~\ref{eq:RTKreads}, when generalized to all operations (i.e., when the predicate holds $\forall op \in H$).
We note that weak fork-linearizability and fork* consistency are incomparable \cite{Cachin.Keidar.ea:11}.

\subsection{Composite and tunable semantics}
\label{subsec:tunable}
To bridge the gap between strongly consistent and efficient implementations, several works have proposed consistency models that entail the use of different semantics in an adaptive fashion 
according to the contingent tradeoffs of performance and correctness.\footnote{We do not formulate formal definitions for tunable semantics considering that they can be expressed by combining the logical predicates reported in the rest of the paper.}

The idea of distinguishing operations' consistency requirements by their semantics dates back to 
the shared-memory systems era.
In that context, consistency models that employed different ordering constraints depending
on operations' types (e.g., \emph{acquire} and \emph{release}, rather than read/write data accesses) 
were called \emph{hybrid}, whereas those that did not operate distinctions were referred to as
\emph{uniform} \cite{Mosberger:93,Dubois.Scheurich.ea:86,Gharachorloo.Lenoski.ea:90}.

A first formal definition which presents a similar diversification was proposed by 
\citeN{Attiya.Friedman:92} for shared-memory multiprocessors.
\textbf{Hybrid} consistency is defined as a model requiring a concerted adoption 
of weak and strong consistency semantics.
In a hybrid consistent system \emph{strong} operations are guaranteed to be seen 
in some sequential order by all processes (as in sequential consistency), 
while \emph{weak} operations are designed to be fast, and they eventually become visible by all processes
(much like in eventual consistency).
Weak operations are only guaranteed to be ordered according to their interleaving with strong operations:
if two operations belong to the same session and one of them is strong, 
then their relative order of invocation is respected and visible by all processes. 
In a similar manner, \citeN{Ladin.ea:92} tackle the tradeoff between performance and consistency 
by assigning to each operation an ordering type.
\emph{Causal} operations respect causality ordering among them,
\emph{forced} operations are delivered in the same order at all relicas, 
and \emph{immediate} operations are performed as they return and they are delivered by each replica in same order with respect to all other operations.

\textbf{Eventual serializability}\footnote{We remark that despite the affinity of its name with
those of popular transactional consistency models, eventual serializability 
has been conceived for non-transactional storage systems.}
is described in \cite{Fekete.Gupta.ea:96} as a condition that requires a partial ordering
of operations which eventually settle to a total order.
According to this model, operations might be \emph{strict} or \emph{non-strict}.
Strict operations are required to be \emph{stable} as soon as they obtain a response, while non-strict ones
may be reordered afterwards.
An operation is said to be \emph{stable} if the prefix of operations preceding it reached a final total order.
\citeN{Fekete.Gupta.ea:96} envision an implementation in which processes issue operations attaching to them both the list of 
identifiers of operations that must be ordered before the requested operation, and a flag that indicates the type of operation
(i.e., strict or non-strict).
The final global and total order achieved by operations can be regarded as a 
sequential consistency ordering as no real-time notion is involved. 

Similarly, \citeN{Serafini.Dobre.ea:10}, distinguish \emph{strong} and \emph{weak} operations.
While strong operations are immediately linearized, weak ones are linearized only eventually.
Weak operations are thus said to respect \textbf{eventual linearizability}.
Weak operations are in fact designed to terminate despite failures,
and can therefore violate linearizability for a finite period of time.
Essentially, eventual linearizability mandates that operations must be ordered 
according to their real-time ordering, yet this applies only to operations invoked after a certain time \emph{t}.
Therefore, earlier operations may have observed inconsistent histories 
and can be temporarily ordered in an arbitrary manner.
Ultimately, the operations in a system that implements eventual linearizability gravitate towards a 
total order that satisfies real-time constraints.

\citeN{Krishnamurthy.Sanders.ea:02}
propose a QoS model that allows client applications of a distributed storage system
to express their consistency requirements. 
According to their requirements, clients are then directed by a middleware towards a 
specific group of replicas implementing synchronous or lazy replication schemes, 
thus applying strong or weak consistency semantics.
This framework is said to provide \textbf{tunable} consistency.

In the same vein, \citeN{Li.Porto.ea:12} propose \textbf{RedBlue} consistency. 
With RedBlue consistency operations are flagged as \emph{blue} or \emph{red}
depending on several conditions such as their commutativity and the respect of invariants.
According to such classification, operations are then 
executed locally and replicated in an eventually consistent manner, 
or serialized with respect to each other through synchronous coordination.
In a follow-up work, \citeN{Li.ea:14} implement and evaluate a system that
would relieve the programmer
from having to choose the right consistency level for each operation
by exploiting a combination of automatic static and dynamic code analysis.

\citeN{Yu.Vahdat:02} propose a continuous consistency spectrum 
based on three metrics: \emph{staleness}, \emph{order error} and \emph{numerical error}.
Those metrics are embedded in a \textbf{conit} (portmanteau of ``consistency unit''), 
which is a three-dimensional vector that quantifies the divergence from an ideal linearizable execution.
Numerical error accounts for the number of write operations that are already 
globally applied but not yet propagated to a given replica of a certain object.
Order error quantifies the number of writes 
at any replica that are subject to reordering, 
while staleness bounds the real-time delay of writes propagation among replicas.
Those metrics are an attempt to capture the semantics of some fundamental dimensions of consistency, 
notably those related to the general requirements of agreement on state and update ordering.
Note that, according to this model, and unlike timed consistency (see Section \ref{subsec:timed}), 
time-based staleness is defined from the replicas' viewpoint rather than 
with respect to the timing of individual operations.

Similarly, \citeN{Santos.Veiga.ea:07} aim at quantifying the 
divergence of data object replicas by using a three-dimensional consistency vector.
Originally designed for distributed multiplayer games on ad-hoc networks,
\textbf{vector-field} consistency mandates for each object a vector 
$\kappa = [\theta, \sigma, \nu ]$ that bounds its 
staleness in a particular \emph{view} of the virtual world.
In particular, the vector establishes the maximum divergence of replicas
in time ($\theta$), number of updates ($\sigma$), and object value ($\nu$).
Unlike conit, this model brings about a notion of locality-awareness as it describes
consistency as a vector field deployed throughout the gaming virtual environment.

Later works put forward tunable consistency as a suitable model for cloud storage,
since it would enables more flexible quality of service (QoS) policies and 
service-level agreements (SLAs). \citeN{Kraska.Hentschel.ea:09} envision consistency \textbf{rationing}, 
which would entail adapting the consistency level at runtime by taking into account economic concerns.
Similarly, \citeN{Chihoub.Ibrahim.ea:12} explore the possibility 
of a self-adaptive protocol that dynamically adjusts consistency to meet the application needs.
In a sequent work, \citeN{Chihoub.Ibrahim.ea:13} add the monetary cost to the equation 
and study its tradeoffs with consistency in cloud settings.
\citeN{Terry.Prabhakaran.ea:13} advocate the use of 
declarative consistency-based SLAs that would allow users of cloud key-value stores to attain
a better awareness of the inherent performance-correctness tensions.
This approach has been subsequently implemented as a declarative programming model 
for tunable consistency by \citeN{Sivaramakrishnan.ea:15}. 

In another attempt at providing stronger consistency semantics for geo-replicated storage, \citeN{Balegas.Duarte.ea:15} introduce \emph{explicit} consistency.
Besides providing eventual consistency, a replicated store implementing explicit consistency ensures that 
application-specific correctness rules (i.e., \emph{invariants}) be respected during executions.
In a follow-up work, \citeN{Gotsman.ea:16} propose a proof rule to help programmers
in the task of assigning fine-grained restrictions on operations in order to respect data integrity invariants. 

Finally, in the context of combining different consistency models, it is worth also mentioning systems that turn eventual consistency of data (provided by modern commodity cloud storage services) into linearizability, by relying on comparably small volumes of metadata stored separately from data in linearizable storage. In independent efforts, this technique was recently proposed under the names of \emph{consistency anchor} \cite{Bessani.Mendes.ea:14} and \emph{consistency hardening} \cite{Dobre.ea:14}.

\subsection{Per-object semantics}
\label{subsec:perobject}
Per-object (or per-key) semantics have been defined to express consistency constraints on a per-object basis. Intuitively, per-object ordering semantics allow for more efficient implementations than global ordering semantics, i.e., across invocations on all objects, taking advantage of techniques such as sharding and state partitioning. 

\textbf{Slow memory}, defined by \citeN{Hutto.Ahamad:90}, is a weaker variant of PRAM consistency.
A shared-memory system implementing this condition requires
that all processes see the writes of a given process to a given object in the same order. In other words, slow memory delineates a per-object weakening of PRAM consistency:
\begin{align}
\textsc{PerObjectPRAM} \triangleq (so \cap ob) \subseteq vis 
\end{align}

An important concept in this family of semantics is that of \textbf{coherence} 
\cite{Gharachorloo.Lenoski.ea:90} (or cache consistency \cite{Goodman:89})
which was first introduced as correctness condition of memory hierarchies in 
shared-memory multiprocessor systems \cite{Dubois.Scheurich.ea:86}.
Coherence ensures that what has been written to a specific memory location becomes visible in some sequential order by all processors, 
possibly through their local caches.
In other words, coherence requires operations to be globally ordered on a per-object basis.
A very similar notion has been coined in recent works \cite{Cooper.Ramakrishnan.ea:08,Lloyd.Freedman.ea:11} 
as \textbf{per-record timeline} consistency.
This condition, described in relation to replicated storage,
ensures that for each individual key (or object), all processes observe the same ordering of operations.
Formally, we capture such condition with the following predicate:
\begin{multline}
\textsc{PerObjectSingleOrder} \triangleq \\ \exists H' \subseteq \{op \in H : op.oval = \nabla\}: 
ar \cap ob = vis \cap ob \setminus(H' \times H) 
\end{multline}

Moreover, a system in which executions respect 
ordering of operations by a certain process on each object and 
a global ordering of all operations invoked on each object,
would implement a semantic condition that we could name as \emph{per-object sequential} consistency:
\begin{multline}
\textsc{PerObjectSequential}(\mathcal{F}) \triangleq \\
\textsc{PerObjectSingleOrder} \wedge \textsc{PerObjectPRAM} \wedge \RVAL
\end{multline}

\textbf{Processor} consistency, defined by \citeN{Goodman:89}
and formalized by \citeN{Ahamad.Bazzi.ea:93}, is expressed by two conditions:
(a) writes issued by a process must be observed in the order in which they were issued, and
(b) if there are two write operations to the same object, 
all processes observe these operations in the same order.
Evidently, the two conditions just mentioned are in fact PRAM and per-record timeline consistency, thus:
\begin{align}
\textsc{ProcessorConsistency}(\mathcal{F}) \triangleq \textsc{PerObjectSingleOrder} \wedge \textsc{PRAM} \wedge \RVAL
\end{align}

In addition, few works in literature (e.g., \cite{Moraru.ea:13}) mention \emph{per-object linearizability},
which is in fact equivalent to linearizability on a per-object basis,
due to its \emph{locality} property \cite{Herlihy.Wing:90}.

We further note that one could compose other arbitrary consistency models by refining some of the
predicates mentioned in this work to match only operations performed on individual objects.
As a case in point, \citeN{Burckhardt.Gotsman.ea:14} 
describe \textbf{per-object causal} consistency as a restriction of causal consistency on a per-object basis,
which leverages the \emph{per-object happens-before} order, defined as: $hbo \triangleq ((so \cap ob) \cup vis)^+$.

\subsection{Synchronized models}
\label{subsec:sync}
For completeness, in this section we overview semantic conditions defined in the '80s 
and early '90s in order to model the correctness of multiprocessor shared-memory systems.
In order to exploit the computational parallelism of these systems, and, at the 
same time, to cope with the different performance of the various components 
(e.g., memories, interconnections, processors, etc.), buffering and caching layers were adopted.
Consequently, the fundamental challenge of this kind of architecture is making sure that 
all memories reflect a common, consistent view of shared data.
Thus, system designers employed \emph{synchronization variables}, i.e.,
special shared objects that only expose two operations, named \emph{acquire} and \emph{release}.
The synchronization variables are used as a generic abstraction for implementing logical fences
meant to control concurrent accesses to shared data objects.
In other words, synchronization variables protect the access to shared data 
through the implementation of mutual exclusion by means of low level primitives 
(e.g., locks) or high-level language constructs (e.g., critical sections).
While the burden of using such tools is left to the programmer, 
the system is supposed to distinguish the accesses to shared data from those to 
the synchronization variables, possibly by implementing and exposing specific low level instructions.

Sequential consistency \cite{Lamport:79} (which we defined in Section \ref{subsec:pram-seq})
was initially adopted as ideal correctness condition for multiprocessors shared-memory systems.
\textbf{Weak ordering}\footnote{Some works in literature refer to weak ordering as to ``weak consistency''. 
We chose to avoid this equivocation by adopting its original nomenclature.}
as described by \citeN{Dubois.Scheurich.ea:86} represents a convenient weakening
of sequential consistency that brings about performance improvements.
In a system that implements weak ordering: 
(a) all accesses to synchronization variables must be \emph{strongly ordered}, 
(b) no access to a synchronization variable is allowed before all previous reads have been completed, and 
(c) processes cannot perform reads before issuing an access to a 
synchronization variable.
In particular, \citeN{Dubois.Scheurich.ea:86} define operations as
strongly ordered if they comply with two specific criteria that constrain 
the ordering of operations according to their session ordering and
relatively to some special instructions supported by pipelined cache-based systems.
Weak ordering has been subsequently redefined in terms of coordination requirements between software and hardware.
Namely, \citeN{Adve.Hill:90} define a \emph{synchronization model}
as a set of constraints on memory accesses that specify how and when synchronization needs to be enforced.
Given this definition,
``\emph{a hardware is weakly ordered with respect to a given synchronization 
model if and only if it appears sequentially consistent to all software that obey the synchronization model}''.

\textbf{Release} consistency, presented by \citeN{Gharachorloo.Lenoski.ea:90}, is 
a weaker extension of weak ordering that exploits further detailed information about synchronization operations 
(i.e., acquire and release) and non-synchronization accesses.
Operations have to be labelled before execution by the programmer (or the compiler) as strong or weak. 
Hence, this widens the classification operated by weak ordering,
which included just synchronization and non-synchronization labels.
Similarly to hybrid consistency (see Section \ref{subsec:tunable}), 
strong operations are ordered according to processor or sequential consistency, 
whereas the ordering of weak operations is just restricted by the relative ordering
with respect to the strong operations invoked by the same process.

Subsequently, several algorithms that slightly alter the original implementation of release consistency
have been designed.
For instance, \textbf{lazy release} consistency \cite{Keleher.ea:92} 
is a relaxed implementation of release consistency
in which actions that enforce consistency are postponed from the release to the next acquire operation.
The rationale of lazy release consistency is reducing the number of 
messages and the amount of data exchanged in a distributed shared-memory system implemented in software.
On the same line, the protocol called \emph{automatic update release consistency} \cite{Iftode.Cezary.ea:96} 
aims at improving performance substantially over software-only implementation of lazy release consistency, 
by using an automatic update mechanism provided by a virtual memory mapped network interface.

\citeN{Bershad.Zekauskas:91} define \textbf{entry} consistency by strengthening the 
relation between synchronization objects and the data which they guard.
According to entry consistency, every object has to be guarded by a synchronization variable.
Thus, in a sense, this model is a location-relative weakening of a consistency semantic, similarly to the models surveyed in Section \ref{subsec:perobject}.
Moreover, entry consistency operates a further distinction of the synchronization operations
in exclusive and non-exclusive.
Thanks to these features, 
reads can occur with a greater degree of concurrency, thus
enabling better performance.

\textbf{Scope} consistency \cite{Iftode.Jaswinder.ea:96} claims 
to offer most of the potential performance advantages
of entry consistency, without requiring explicit binding of data to synchronization variables.
The key intuition of scope consistency is the use of an abstraction called \emph{scope} to implicitly capture the relationship
between data and synchronization operations.
Consistency scopes can be derived automatically from the use of synchronization variables in the program,
thus easing the work of programmers.

With the definition of \textbf{location} consistency, \citeN{Gao.Sarkar:00} forwent the basic 
assumption of \emph{memory coherence} \cite{Gharachorloo.Lenoski.ea:90}, 
i.e., the property that ensures that all writes to the same object are observed in the same order by all processes (see Section \ref{subsec:perobject}).
Thus, they explored the possibility of executing multithreaded programs in a correct manner by just 
exploiting a partial order on writes to shared data.
Similarly to entry consistency, in location consistency each object is associated to 
a synchronization variable. However, thanks to the relaxed undelying ordering constraint, 
\citeN{Gao.Sarkar:00} prove that location consistency can be more efficient and equivalently strong 
when it is applied to settings with low data contention between processes.

\section{Related work}
\label{sec:relwork}
Several works in literature have provided overviews on consistency models.
In this section we classify these works according to their different perspectives.

\paragraph*{Shared-memory systems}
\citeN{Gharachorloo.Lenoski.ea:90} proposed
a classification of shared memory access policies, 
specifically regarding their concurrency control semantics 
(e.g., synchronization operations versus read/write accesses). 
\citeN{Mosberger:93} adopted this classification to 
conduct a study on the memory consistency models popular at that time
and their implementation tradeoffs.
\citeN{Adve.Gharachorloo:96} summarized in a practical tutorial 
the informal definitions and related issues of consistency models most commonly adopted 
in shared-memory multiprocessor systems.

Several subsequent works developed uniform frameworks and notations
to represent consistency semantics defined in literature \cite{Adve.Hill:93,Raynal.Schiper:97,Bataller.Bernabeu:97}.
Most notably, \citeN{Steinke.Nutt:04} provide a unified theory of consistency models
for shared memory systems based on the composition of few fundamental declarative properties.
In turn, this declarative and compositional approach outlines a partial ordering over consistency semantics.
Similarly, a treatment of composability of consistency conditions had been carried out in \cite{Friedman.ea:03}.

While all these works proved to be valuable and formally sound,
they represent only a limited portion of the consistency semantics
relevant to modern non-transactional storage systems.

\paragraph*{Distributed storage systems}
In more recent years, researchers have been proposing categorizations of the most influential
consistency models for modern storage systems.
Namely, \citeN{Tanenbaum.Steen:07} proposed the client-centric versus
data-centric classification, while \citeN{Bermbach.Kuhlenkamp:13},
expanded such classification and provided descriptions for the most popular models.
While practical and instrumental in attaining a good understanding of the consistency spectrum,
these works propose informal treatments based on a simple dichotomous categorization
which falls short of capturing some important consistency semantics.
With this survey we aim at improving over these works, as we adopt a formal model
based on first-order logic predicates and graph theory.
We derived this model from the one proposed in \cite{Burckhardt:14}, 
which we modified and expanded in order to enable the definition of a wider and richer range of consistency semantics. Moreover, whereas \citeN{Burckhardt:14} focuses mostly on session and eventual semantics, we cover a broader ground including more than 50 different consistency semantics. 

\paragraph*{Measuring consistency}
A concurrent research trend has been straining to design uniform and 
rigorous frameworks to measure consistency in both shared memory systems 
and, more recently, in distributed storage systems.
Namely, while some works have proposed metrics to assess consistency \cite{Yu.Vahdat:02,Golab.ea:14},
others have devised methods to verify, given an execution, whether it satisfies a certain consistency model \cite{Misra:86,Gibbons.ea:97,Anderson.Li.ea:10}.
Finally, due to the loose definitions and opaque implementations of eventual consistency, 
recent research has tried to quantify its inherent anomalies as perceived from a client-side perspective
\cite{Wada.Fekete.ea:11,Patil.ea:11,Bermbach.Tai:11,Muntasir.Wojciech.ea:2012,Lu.ea:15}.
In this regard, our work provides a more comprehensive and structured overview of the metrics 
that can be adopted to evaluate consistency.

\paragraph*{Transactional systems}
Readers interested in pursuing a formal treatment of the most important
consistency models for transactional storage systems may refer to \cite{Adya:99}.
Similarly, other works by \citeN{Harris.ea:10} and by \citeN{Dziuma.Fatourou.ea:14}
complement this survey with overviews on models specifically designed for transactional memory systems.
Finally, some recent research \cite{Burckhardt.Leijen.ea:12,Cerone.Bernardi.ea:15}
adopted variants of the same framework used in this paper to propose 
axiomatic specifications of transactional consistency models.

\section{Conclusion}
\label{sec:conclusion}

In this work we presented an overview of the most relevant
consistency models for non-transactional storage systems.
Thanks to our methodical approach,
we were able to highlight subtle yet meaningful differences among 
consistency models, thus helping scholars and practitioners attain a
better understanding of the tradeoffs involved.

To describe consistency semantics we adopted a mathematical framework 
based on graph theory and first-order logic.
As first contribution of this work, we developed such formal framework as an extension 
of the one presented in \cite{Burckhardt:14}.
The framework is comprehensive and useful in capturing different factors involved in the executions of a  distributed storage system.

We used this framework to formulate formal definitions for the most popular of the over 50 consistency semantics we analyzed.
For the rest of them, we presented informal descriptions which provide insights about their feature and  relative strenghts.
Moreover, thanks to the axiomatic approach we adopted, we laid out a clustering of semantics according to criteria which account for their natures and common traits.
In turn, both the clustering and the formal definitions helped us building a partial ordering of consistency models (see Figure~\ref{fig:non-trans_graph}).
We believe this partial ordering of semantics will prove convenient both in designing more precise 
and coherent models, and in evaluating and comparing the correctness of systems already in place.
Finally, as further contribution, we provide in Appendix~\ref{sec:reftable} an ordered list of all the models analyzed in this work, along with references to 
their definitions and main implementations in research literature.

\section*{Acknowledgement}
We would like to thank Alysson Bessani, Christian Cachin, Marc Shapiro, and
the anonymous reviewers for their helpful comments on this work.
This research was supported in part by the EU projects
CloudSpaces (FP7-317555) and SUPERCLOUD (Horizon 2020 programme, grant No. 643964).

\bibliographystyle{ACM-Reference-Format-Journals}
\bibliography{consistency}

\clearpage
\appendix
\section{Summary of consistency predicates}
\label{sec:predicates}
\renewcommand{\arraystretch}{1.5}

\begin{flushleft}
\small
	\begin{longtabu} to \linewidth {X[4,l] | X[7.5,l]}
		$\textsc{Linearizability}(\mathcal{F})$ & $\textsc{SingleOrder} \wedge \textsc{RealTime} \wedge \RVAL$ \\
		$\textsc{SingleOrder}$ & $\exists H' \subseteq \{op \in H : op.oval = \nabla\}: vis = ar \setminus(H' \times H)$\\
		$\textsc{RealTime}$ & $rb \subseteq ar$ \\
		$\textsc{Regular}(\mathcal{F})$ & $\textsc{SingleOrder} \wedge \textsc{RealTimeWrites} \wedge \RVAL$ \\
		$\textsc{Safe}(\mathcal{F})$ & $\textsc{SingleOrder} \wedge \textsc{RealTimeWrites} \wedge \textsc{Seq}\RVAL$ \\
		$\textsc{RealTimeWrites}$ & $rb|_{wr \rightarrow op} \subseteq ar$ \\
		$\textsc{Seq}\RVAL$ & $\forall op \in H : \mathit{Concur(op)}=\emptyset \Rightarrow op.oval \in \mathcal{F}(op, cxt(\A,op))$ \\
		
		$\textsc{EventualConsistency}(\mathcal{F})$ & $\textsc{EventualVisibility} \wedge \textsc{NoCircularCausality} \wedge \RVAL$ \\
		$\textsc{EventualVisibility}$ & $\forall a \in H, \forall [f] \in H/\approx_{ss} : |\{ b \in [f] : (a \xrightarrow[]{\text{\textit{rb}}} b) \wedge (a \xnrightarrow{vis} b)\}| < \infty $ \\
		$\textsc{NoCircularCausality}$ & $acyclic(hb)$ \\
		$\textsc{StrongConvergence}$ & $\forall a,b\in H|_{rd}: vis^{-1}(a)|_{wr}=vis^{-1}(b)|_{wr} \Rightarrow a.oval = b.oval$\\
		$\textsc{StrongEventualCons.}(\mathcal{F})$ & $\textsc{EventualConsistency}(\mathcal{F}) \wedge \textsc{StrongConvergence}$ \\
		$\textsc{QuiescentConsistency}(\mathcal{F})$ & $|H|_{wr}| < \infty \Rightarrow \exists C \in \mathcal{C} : \forall[f] \in H/\approx_{ss} : |\{op \in [f]: op.oval \notin \mathcal{F}(op,C) \}| < \infty  $ \\
		
		$\textsc{PRAM}$ & $so \subseteq vis$ \\
		$\textsc{SequentialConsistency}(\mathcal{F})$ & $\textsc{SingleOrder} \wedge \textsc{PRAMConsistency} \wedge \RVAL$\\
		
		$\textsc{MonotonicReads}$ & $\forall a\in H, \forall b,c \in H|_{rd} : a \xrightarrow[]{\text{vis}} b \wedge b \xrightarrow[]{\text{so}} c \Rightarrow a \xrightarrow[]{\text{vis}} c \triangleq (vis;so|_{rd \rightarrow rd}) \subseteq vis$ \\
		$\textsc{ReadYourWrites}$ & $\forall a\in H|_{wr}, \forall b \in H|_{rd} : a \xrightarrow[]{\text{so}} b \Rightarrow a \xrightarrow[]{\text{vis}} b \triangleq so|_{wr \rightarrow rd} \subseteq vis$ \\
		$\textsc{MonotonicWrites}$ & $\forall a, b \in H|_{wr} : a \xrightarrow[]{\text{so}} b \Rightarrow a \xrightarrow[]{\text{ar}} b \triangleq so|_{wr \rightarrow wr} \subseteq ar$ \\
		$\textsc{WritesFollowReads}$ & $\forall a,c\in H|_{wr}, \forall b\in H|_{rd} : a \xrightarrow[]{\text{vis}} b \wedge b \xrightarrow[]{\text{so}} c \Rightarrow a \xrightarrow[]{\text{ar}} c \triangleq (vis;so|_{rd \rightarrow wr}) \subseteq ar$ \\
		
		$\textsc{CausalVisibility}$ & $hb \subseteq vis$ \\
        $\textsc{CausalArbitration}$ & $hb \subseteq ar$ \\
		$\textsc{Causality}(\mathcal{F})$ & $\textsc{CausalVisibility} \wedge \textsc{CausalArbitration} \wedge \RVAL$ \\
		$\textsc{Causal+}(\mathcal{F})$ & $\textsc{Causality}(\mathcal{F}) \wedge \textsc{StrongConvergence}$ \\
			$\textsc{RealTimeCausality}(\mathcal{F})$ & $\textsc{Causality}(\mathcal{F}) \wedge \textsc{RealTime}$ \\
		
		$\textsc{TimedVisibility}(\Delta)$ &  $\forall a\in H|_{wr},\forall b \in H, \forall t \in \mathit{Time} : a.rtime = t \wedge b.stime = t + \Delta \Rightarrow a \xrightarrow[]{\text{vis}} b$ \\
        $\textsc{TimedCausality}(\mathcal{F},\Delta)$ & $\textsc{Causality}(\mathcal{F}) \wedge \textsc{TimedVisibility}(\Delta)$ \\
        $\textsc{TimedLinearizability}(\mathcal{F},\Delta)$ & $\textsc{SingleOrder} \wedge \textsc{TimedVisibility}(\Delta) \wedge \RVAL$\\        
        $\textsc{PrefixSequential}(\mathcal{F})$ & $\textsc{SingleOrder} \wedge \textsc{MonotonicWrites} \wedge \RVAL$ \\
        $\textsc{PrefixLinearizable}(\mathcal{F})$ & $\textsc{SingleOrder} \wedge \textsc{RealTimeWW} \wedge \RVAL$ \\
        $\textsc{RealTimeWW}$ & $rb|_{wr \rightarrow wr} \subseteq ar$\\
        $\textsc{K-Linearizable}(\mathcal{F},K)$ & $ \textsc{SingleOrder} \wedge \textsc{RealTimeWW} \wedge \textsc{K-RealTimeReads}(K) \wedge \RVAL$ \\
        $\textsc{K-RealTimeReads}(K)$& $\forall a \in H|_{wr}, \forall b\in H|_{rd}, \forall PW \subseteq H|_{wr}, \forall pw\in PW: |PW|<K \wedge a\xrightarrow[]{\text{\textit{ar}}}pw \wedge pw\xrightarrow[]{\text{\textit{rb}}}b \wedge a\xrightarrow[]{\text{\textit{rb}}}b \Rightarrow a\xrightarrow[]{\text{\textit{ar}}}b$\\        
        
        $\textsc{ForkLinearizability}(\mathcal{F})$ & $\textsc{PRAM} \wedge \textsc{RealTime} \wedge \textsc{NoJoin} \wedge{\RVAL} $\\
        $\textsc{NoJoin}$ & $\forall a_i,b_i,a_j,b_j\in H: 
        a_i \not\approx_{ss} a_j \wedge 
        (a_i,a_j)\in ar\setminus vis \wedge a_i \preceq_{so} b_i \wedge a_j \preceq_{so} b_j   \Rightarrow 
        (b_i,b_j),(b_j,b_i)\notin vis$ \\
        $\textsc{Fork*}(\mathcal{F})$ & $\textsc{ReadYourWrites} \wedge \textsc{RealTime} \wedge \textsc{AtMostOneJoin} \wedge{\RVAL} $\\
		$\textsc{AtMostOneJoin}$ & $\forall a_i,a_j\in H: 
		a_i \not\approx_{ss} a_j \wedge  (a_i,a_j)\in ar\setminus vis \Rightarrow
		|\{ b_i \in H: a_i \preceq_{so} b_i \wedge 
		(\exists b_j\in H: a_j \preceq_{so} b_j \wedge b_i\xrightarrow[]{\text{\textit{vis}}}b_j\}| \le 1 
		\wedge |\{ b_j \in H: a_j \preceq_{so} b_j \wedge 
		(\exists b_i\in H: a_i \preceq_{so} b_i \wedge b_j\xrightarrow[]{\text{\textit{vis}}}b_i\}| \le 1$\\
        $\textsc{ForkSequential}(\mathcal{F})$ & $\textsc{PRAM} \wedge \textsc{NoJoin} \wedge{\RVAL}$ \\
        $\textsc{WeakForkLin}(\mathcal{F})$ & $\textsc{PRAM} \wedge \textsc{K-RealTime(2)} \wedge \textsc{AtMostOneJoin} \wedge{\RVAL}$ \\
        
        $\textsc{PerObjectPRAM}$ & $(so \cap ob) \subseteq vis$ \\
        $\textsc{PerObjectSingleOrder}$ & $\exists H' \subseteq \{op \in H : op.oval = \nabla\}: 
        ar \cap ob = vis \cap ob \setminus(H' \times H) $ \\
        $\textsc{PerObjectSequential}(\mathcal{F})$ & $\textsc{PerObjectSingleOrder} \wedge \textsc{PerObjectPRAM} \wedge{\RVAL}$\\
        $\textsc{ProcessorConsistency}(\mathcal{F})$ & $\textsc{PerObjectSingleOrder} \wedge \textsc{PRAM} \wedge{\RVAL}$\\
        $\textsc{PerObjectHappensBefore}$ & $hbo \triangleq ((so \cap ob) \cup vis)^+$ \\
	\end{longtabu}
	\captionof{table}{Summary of consistency predicates listed in the paper.}
	\label{tab:models}
\end{flushleft}

\clearpage
\section{Primary references}
\label{sec:reftable}
\renewcommand{\arraystretch}{1.11}
\begin{center}
	\begin{longtabu}to \textwidth {
                        X[1,l]|
                        X[1.5,l]|
                        X[1.5,l]}
	\hline
		Models           & Definitions              & Implementations\footnote{In case of very popular consistency semantics (e.g., causal consistency, atomicity/linearizability), we only cite a subset of known implementations.} \\
	\hline

		Atomicity       & \cite{Lamport:86:vol2}                & \cite{Attiya.ea:95} \\
		Bounded fork-join causal         & \cite{P-Mahajan.Dahlin:11}                & - \\
		Bounded staleness         & \cite{Mahajan.Setty.ea:10}                & - \\
		Causal         & \cite{Lamport:78,Hutto.Ahamad:90,Ahamad.Neiger.ea:95,P-Mahajan.Dahlin:11}                & \cite{Ladin.ea:92,Birman.ea:91,Lakshmanan.ea:01,Lloyd.Freedma:13,Du.ea:14,Zawirski.ea:15,Lesani.Bell.ea:16} \\ 		Causal+         & \cite{Lloyd.Freedman.ea:11}                & \cite{Petersen.ea:97,Belaramani.ea:06,Almeida.Leitao.ea:13} \\
		Coherence         & \cite{Dubois.Scheurich.ea:86}                & - \\
		Conit         & \cite{Yu.Vahdat:02}                & - \\
				$\Gamma$-atomicity         & \cite{Golab.ea:14}     & - \\ 
		$\Delta$-atomicity         & \cite{Golab.Li.ea:11}                & - \\
		Delta         & \cite{Singla.ea:97}                & - \\
		Entry         & \cite{Bershad.Zekauskas:91}                & - \\
		Eventual        & \cite{Terry.Demers.ea:94,Vogels:08}     & \cite{Reiher.ea:94,DeCandia.Hastorun.ea:07,Singh.ea:09,Bortnikov.Chockler.ea:10,Bronson.ea:13} \\ 		Eventual linearizability         & \cite{Serafini.Dobre.ea:10}                & - \\
		Eventual serializability         & \cite{Fekete.Gupta.ea:96}                & - \\
		Fork*         & \cite{Li.Mazieres:07}                & \cite{Feldman.Zeller.ea:10} \\
		Fork         & \cite{Mazieres.Shasha:02,Cachin.ea:07} & \cite{Li.Krohn.ea:04,Brandenburger.Cachin.ea:15} \\
		Fork-join causal         & \cite{Mahajan.Setty.ea:10}                & - \\
		Fork-sequential         & \cite{Oprea.Reiter:06}                & - \\
		Hybrid         & \cite{Attiya.Friedman:92}                & - \\
		K-atomic         & \cite{Aiyer.ea:05}                & - \\
		K-regular         & \cite{Aiyer.ea:05}                & - \\
		K-safe         & \cite{Aiyer.ea:05}                & - \\
		$k$-staleness         & \cite{Bailis.Venkataraman.ea:12}                & - \\
		Lazy release         & \cite{Keleher.ea:92}                & - \\
        Linearizability & \cite{Herlihy.Wing:90}                & \cite{Burrows:06,Baker.Bond.ea:11,Glendenning.Beschastnikh.ea:11,Calder.Wang.ea:11,Corbett.Dean.ea:13,Han.Shen.ea:15,Lee.ea:15} \\
		Location         & \cite{Gao.Sarkar:00}                & - \\
		Monotonic reads         & \cite{Terry.Demers.ea:94}                & \cite{Terry.ea:95} \\
		Monotonic writes        & \cite{Terry.Demers.ea:94}                & \cite{Terry.ea:95} \\
		Observable causal       & \cite{Attiya.ea:15} & - \\
		PBS $\langle$\emph{k, t}$\rangle$-staleness         & \cite{Bailis.Venkataraman.ea:12}                & - \\		
		Per-object causal & \cite{Burckhardt.Gotsman.ea:14} & - \\
				Per-record timeline         & \cite{Cooper.Ramakrishnan.ea:08,Lloyd.Freedman.ea:11}                & \cite{Andersen.ea:09} \\ 		PRAM            & \cite{Lipton.Sandberg:88}                & - \\
		Prefix         & \cite{Terry.ea:95,Terry:13}                 & - \\
		Processor         & \cite{Goodman:89}                & - \\
		Quiescent       & \cite{Herlihy.Shavit:08}                & - \\
		Rationing         & \cite{Kraska.Hentschel.ea:09}                & - \\
		Read-your-writes        & \cite{Terry.Demers.ea:94}                & \cite{Terry.ea:95} \\
		Real-time causal         & \cite{P-Mahajan.Dahlin:11}                 & - \\
		RedBlue         & \cite{Li.Porto.ea:12}                & - \\
		Regular         & \cite{Lamport:86:vol2}                & \cite{Malkhi.Reiter:98b,Guerraoui.Vukolic:06} \\
		Release         & \cite{Gharachorloo.Lenoski.ea:90}                & - \\
		Safe            & \cite{Lamport:86:vol2}                & \cite{Malkhi.Reiter:98,Guerraoui.Vukolic:06} \\
		Scope         & \cite{Iftode.Jaswinder.ea:96}                & - \\
		Sequential              & \cite{Lamport:79}                & \cite{Rao.Shekita.ea:11} \\
		Slow         & \cite{Hutto.Ahamad:90}                & - \\
		Strong eventual & \cite{Shapiro.ea:11} & \cite{Shapiro.Preguica.ea:11,Conway.Marczak.ea:12,Roh.Jeon.ea:11} \\ 
		Timed causal         & \cite{Torres-Rojas.Meneses.05}                & - \\
		Timed serial         & \cite{Torres-Rojas.Ahamad.ea:99}                & - \\
		Timeline         & \cite{Cooper.Ramakrishnan.ea:08}                & \cite{Rao.Shekita.ea:11} \\ 		Tunable         & \cite{Krishnamurthy.Sanders.ea:02}                & \cite{Lakshman.Malik:10,Wu.Butkiewicz.ea:13,Perkins.Agrawal.ea:15,Sivaramakrishnan.ea:15} \\
		                                                    		$t$-visibility         & \cite{Bailis.Venkataraman.ea:12}                & - \\
		Vector-field         & \cite{Santos.Veiga.ea:07}                & - \\
		Weak            & \cite{Vogels:08,Bermbach.Kuhlenkamp:13} & - \\
		Weak fork-linearizability         & \cite{Cachin.Keidar.ea:11}                & \cite{Shraer.ea:10} \\
		Weak ordering         & \cite{Dubois.Scheurich.ea:86}                & - \\
		Writes-follow-reads     & \cite{Terry.Demers.ea:94}                & \cite{Terry.ea:95} \\
	\end{longtabu}
	\captionof{table}{Definitions of consistency semantics and some of their implementations in literature.}

\label{tab:papers}
\end{center}

\end{document}